\documentclass[iop]{emulateapj}
\usepackage{hyperref,mathrsfs,wasysym}  

\font\smc=cmcsc10

\slugcomment{To appear in the Astrophysical Journal, October 2010}

\shorttitle{Magnetic Activity and Rotation Period of $\alpha$ Cen B}
\shortauthors{DeWarf et al.}

\received{2009 December 21; Accepted 2010 August 11}
\begin{document}

\title{X-Ray, FUV, and UV Observations of $\alpha$ Centauri B: \\
Determination of Long-term Magnetic Activity Cycle and Rotation Period}

\author{
          L.E. DeWarf\,\altaffilmark{1},
          K.M. Datin,
          and E.F. Guinan\altaffilmark{1}
        }

\affil{Department of Astronomy and Astrophysics, Villanova University, 
800 Lancaster Avenue, Villanova, PA 19085, USA}

\email{Laurence.DeWarf@Villanova.edu}

\altaffiltext{1}{\,Based on observations made with the NASA-CNES-CSA 
{\it Far Ultraviolet Spectroscopic Explorer (FUSE)}. {\it FUSE} is operated 
for NASA by the Johns Hopkins University under NASA contract NAS5-32985.}

\begin{abstract}

Over the last couple of decades we have been carrying out a study of stellar 
magnetic activity, dynamos, atmospheric physics, and spectral irradiances 
from a sample of solar-type G0-5 V stars with different ages.  One of the 
major goals of this program is to study the evolution of the Sun's X-ray 
through NUV spectral irradiances with age.  Of particular interest is the 
determination of the young Sun's elevated levels of high-energy fluxes 
because of the critical roles that X-ray (coronal) through FUV (transition 
region (TR), chromospheric) emissions play on the photochemical and 
photoionization evolution (and possible erosion) of early, young planetary 
atmospheres and ionospheres.  Motivated by the current exoplanetary search 
missions (such as {\it Kepler} and {\it CoRoT}, along with the planned 
{\it Space Interferometry mission} and {\it Darwin/Terrestrial Planet Finder} 
missions) that are hunting for earth-size planets in the habitable zones 
(liquid water) of nearby main-sequence G-M stars, we are expanding our 
program to cooler, less luminous, but very importantly, much more numerous 
main-sequence K-type stars, such as $\alpha$ Centauri B.  The long life 
(2-3$\times$ longer than our Sun) and slow evolution of K stars provide nearly 
constant energy sources for possible hosted planets.  This program 
parallels our ``Sun in Time'' program, but extends the study to stars with 
deeper convective zone depths.  Presented here are X-ray (coronal; 
{\it ROSAT}, {\it Chandra}, {\it XMM}), UV (TR; {\it IUE}), NUV 
(chromospheric; {\it IUE}), and recently acquired FUV (TR/chromospheric; 
{\it FUSE} Cycles 7/8) observations of the K1 V star $\alpha$ Cen B 
(HD 128621; {\it V} = 1.33; {\it (B-V)} = +0.88; $\tau$ = 5.6 $\pm$ 0.6 Gyr).  
These combined high-energy measures provide a more complete look into the 
nature of $\alpha$ Cen B's magnetic activity and X-UV radiances.  We find that 
$\alpha$ Cen B has exhibited significant long-term variability in X-ray 
through NUV emission fluxes, indicating a solar-like long-term activity cycle 
of {\it P}$_{\rm cycle}$ = 8.84$\pm$0.4 years.  In addition, analysis of the 
short-term rotational modulation of mean light due to the effects of 
magnetically active regions has yielded a well-determined rotation period of 
{\it P}$_{\rm rotation}$ = 36.2$\pm$1.4 days.  $\alpha$ Cen B is the only old 
main-sequence K star with a reliably determined age and rotation period, and 
for early K-stars, as in the case of the Sun for G2 V stars, is an important 
calibrator for stellar age/rotation/activity relations.

\end{abstract}



\clearpage

\section{INTRODUCTION}

\subsection{The ``Sun in Time'' Program}

Since 1990 we have been carrying out an in-depth study of the evolution 
of the chromospheres, transition regions (TRs), and coronae of $\sim$1.0 
{\it M}$_{\odot}$ stars throughout their main-sequence lifetimes (see 
\citealt{DG94a,DG94b,DGD94,GGS97,GGS98}; and more recently, 
\citealt{GRH03,RGGA05,GED09}).  This program, called ``The Sun in Time,'' 
is a comprehensive study across the electromagnetic spectrum using 
a homogeneous sample of single, nearby G0-5 main-sequence stars with 
known rotation periods and well-determined physical properties (viz. mass, 
radius, temperature, etc.)  These stars are used as proxies for the 
Sun at different ages, and thereby help to quantify the effects of spin-down 
due to magnetic braking.  This investigation also bears on the crucial 
question of the influence of the young Sun's strong X-ray and FUV emissions 
on developing planetary systems -- particularly on the photochemical and 
photoionization evolution of early planetary atmospheres.  To this end, we 
have constructed spectral irradiance tables for the Sun at different ages 
(see \citealt{RGGA05}).  These data sets are of great interest to 
researchers in paleo-planetary atmospheres as well as for studies of the 
evolution of the atmospheres for the increasing number (400+ reported as of 
2009 December) of planets now found orbiting other stars.  For example, we 
have recently collaborated with an astrobiology group to study the effect of 
the young Sun's strong X-ray and ultraviolet (XUV) irradiance on the loss of 
water from Mars and its implications for the oxidation of the Martian soil 
\citep{LLKRGAB03}.  In other studies \citep{LSRGB03,GSPLSRGMBW04}, our solar 
proxy data have been used to investigate the atmospheric loss of exoplanets 
resulting from XUV heating, which can eventually lead to the evaporation of 
``hot Jupiters.''  \citet{KLLTRKLLGBB06} have used these data to investigate 
the early evolution and erosion of Venus' atmosphere. \\

\subsection{Expanding to the dK Stars}

More recently, we have expanded this ``Sun in Time'' project to include 
the much more numerous early K-type stars ($\sim$3$\times$ higher space 
density than G-type stars).  The slower evolution of dK stars (due to 
their lower mass and slower nuclear reaction rates) makes these attractive 
stars for hosting habitable planets with long-lived, stable climates.  The 
focus of the overall investigation is twofold -- (1) Modeling of dwarf 
K-type stars to better understand magnetic activity and magnetic energy 
generation (i.e., dynamo), and (2) Constructing complete irradiance 
tables covering the main-sequence evolution of low-mass K-type stars.  Similar 
to the solar proxies, the stars in this program have been selected in a narrow 
spectral type interval (K0-K5 V) and cover a wide range of rotation periods 
(aka ages).  As shown in Table \ref{tab:dKStars}, our sample of targets covers 
ages from $\sim$50 Myr ($P_{\rm rotation}$$\approx$\,0.5 days) to 8-12 Gyr 
($P_{\rm rotation}$$\approx$\,50 days).  These program stars have well 
determined parallaxes, colors, spectral types, and also have observations of 
age-sensitive measures (rotation) such as {\it L}$_{\rm X}$, \ion{C}{4} 
($\lambda$1550\AA), and \ion{Mg}{2} {\it h+k} ($\sim$$\lambda$2800\AA) 
emission fluxes.

This larger study, besides improving our understanding of magnetic 
dynamo-related phenomena, will help to identify and characterize stars that 
might be suitable for life.  dK stars are excellent hosts for habitable 
planets -- possessing long life, stability, and slow changes in luminosity 
with time.  Their habitable zones (HZs) are fixed for billions of years.  
Because of the relatively large number of main-sequence K-type stars 
($\sim$13\% of all stellar types) in the solar neighborhood, and the closeness 
of their habitable regions (0.4 - 0.8 AU), these stars will likely 
be the main targets of exoplanet search missions such as {\it Kepler} and 
{\it CoRoT}, along with the proposed {\it Space Interferometry mission} and 
{\it Darwin/Terrestrial Planet Finder} missions.  An additional important 
attribute of K stars in terms of habitability is that they evolve very slowly 
and have lifetimes from 2 to 5 times that of the Sun.  These investigations 
will also have a major impact on studies of X-ray through NUV radiation and 
their effects on the environments of extrasolar planets and possible origin 
and evolution of extraterrestrial life.

Long-term chromospheric magnetic activity cycle modulations have been 
optically measured for nearly 100 K stars (see, for example, 
\citealt{BDSHFWBRWZBBCDFLMMNPPRRSSVW95}).  At this point, however, the coronal 
X-ray magnetic activity cycles of both G- and K-type stars are largely 
unknown, with only about three reasonable determinations (see 
\citealt{FMOSSH08}).  Presented here is the analysis of the nearest K star, 
$\alpha$ Centauri B.

\section{THE $\alpha$ CENTAURI SYSTEM}

\begin{figure}
\epsscale{0.8}
\plotone{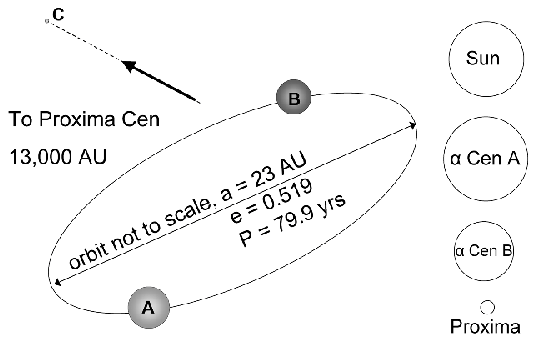}
\caption{Schematic of the $\alpha$ Cen ABC system is shown above (see also 
Table \ref{tab:aCenProp}).  The outlines on the right show the Sun and 
$\alpha$ Cen's components to scale.  The primary component ($\alpha$ Cen A) is 
approximately 9\% more massive and 22\% larger than our Sun, whereas the 
secondary (B) is about 10\% less massive and 14\% smaller.  $\alpha$ Cen B is 
expected to have a deeper and more compact convection zone (CZ $\approx$ 0.5 
{\it R}$_{*}$) and is slightly older ($\tau$ $\approx$ 5-6 Gyr) than our Sun. 
\label{fig:aCen}}
\end{figure}

The $\alpha$ Centauri triple star system is the closest star system to the Sun 
at a distance of 1.347$\pm$0.003 pc \citep{H97,vL07,S99, PNN99}.  $\alpha$ Cen 
AB is a well-separated, tidally non-interacting binary system 
({\it P}$^{\rm \,AB}_{\rm orbit}$ = 79.9 yr; separation $\approx$ 23 AU), 
consisting of a G2 V (A) and a K1 V (B) star.  A third component (C), the dM5e 
star Proxima Centauri, is most likely gravitationally bound to $\alpha$ Cen 
AB, but in a wide orbit separated by about 13,000 AU (see Figure 
\ref{fig:aCen} and Table \ref{tab:aCenProp}).  The secondary ($\alpha$ Cen B; 
HD 128621) is slightly smaller than our Sun with regard to its mass 
({\it M}/{\it M}$_{\odot}$ = 0.90; 
\citealt{DGA86,PNMBTMJPCBPHSRK02,Y08}), radius ({\it R}/{\it R}$_{\odot}$ = 
0.86; \citealt{KTSBLMP03}), and temperature ({\it T}/{\it T}$_{\odot}$ = 
0.92; \citealt{NM97,MPLTB00,PLK08}).  However, $\alpha$ Cen B is expected to 
have a deeper convection zone (CZ $\approx$ 0.5 {\it R}$_{*}$) and is slightly 
older ($\tau$ $\approx$ 5-6 Gyr; see 
\citealt{FA78,DGA86,GD00,TPMBBC02,ECTMMCB04,Y07,Y08}) than our Sun.

Due to its proximity, the $\alpha$ Cen system has been extensively 
researched.  In studies of stellar magnetic activity, $\alpha$ Cen forms an 
astrophysical laboratory to investigate angular momentum loss, coronal and 
chromospheric activity, for three coeval stars with differing masses (1.09, 
0.90, and 0.12 {\it M}$_{\odot}$) and deepening outer CZs (0.3, 0.5, and 
$\sim$1.0 {\it R}$_{*}$).  Also, $\alpha$ Cen B is the oldest K-type 
star with a well-determined age -- determined from the isochronal age of its 
close companion, $\alpha$ Cen A (5-6 Gyr; \citealt{FA78}).  This makes 
$\alpha$ Cen B, along with $\alpha$ Cen A and $\alpha$ Cen C (Proxima), 
crucial calibrators for any age/activity/rotation studies.

The frequency of planets occurring in multiple star systems has recently been 
investigated by \citet{BD07} and \citet{QALC07}.  In particular, the 
possibility of planets hosted in the $\alpha$ Cen AB system has been addressed 
dynamically by \citet{B88}, \citet{WH97}, and \citet{GRDLQF08}.  In fact, 
\citet{TMS09} show that $\alpha$ Cen B may be both plausible, and well-suited 
observationally, for the potential detection of a terrestrial-type planet 
residing within its habitable zone (HZ $\approx$ 0.6 - 1.2 AU).  
Unfortunately, to date there is no evidence for planets hosted by the system 
(see, also, \citealt{MHC93,HKCDD96,EKEHC01}).  However, the $\alpha$ Cen 
AB system has an $\sim$2$\times$ higher metal abundance than our Sun ([Fe/H] 
$\approx$ 0.25; \citealt{FM90,CFCB92,NM97,DORB05,PLK08}).  This may favor the 
formation of terrestrial planets.  \citet{G00} and \citet{WCG07}, for example, 
have shown that extrasolar planets tend to be hosted by stars with higher than 
average metal abundances.

High precision astrometry of Proxima Centauri ($\alpha$ Cen C) carried 
out by \citet{B08} with the fine guidance sensor (FGS) camera aboard the 
{\it Hubble Space Telescope} ({\it HST}), reveals no evidence of 
hosted planets with masses greater than $\sim$5 {\it M}$_{\oplus}$ (see also 
\citealt{BMCNJALCSHFWDSWF99}; \citealt{BMB06}; \citealt{BMB08}).  However, 
this study indicated an 83.1 day variation in brightness, most likely 
originating from the rotational light modulation from starspots.  Analysis of 
over seven years of photometry of Proxima confirms this rotation period 
(see S.G.\citealt{EGM09}, private communication) and also indicates a possible 
$\sim$7 yr starspot cycle.  High-precision radial velocity studies have also 
yielded only upper limits, with {\it m}\,sin\,{\it i} $\le$ 1 M$_{\neptune}$ 
and separations {\it a} $\ge$ 1 AU \citep{KHCDDE99,EK08}.

Although the primary focus of this paper is $\alpha$ Cen B, we now have 
sufficient data to compare the rotation and activity cycles of all three stars 
of this system.  This comparison is intriguing because all three stars are 
coeval, differ mainly in mass, {\it T}$_{\rm eff}$, and convective zone depth.  
Thus the $\alpha$ Cen triple system can serve as a ``mini'' laboratory for 
studying angular momentum loss, activity, and cycle properties.  As part of 
our ongoing research, the rotation periods of $\alpha$ Cen A, B, C are 
observed to be $\sim$15-20, 36.2$\pm$1.4, and 83.1$\pm$0.6 days, 
respectively.  As seen, there is a strong dependence of rotation on mass and 
convective zone depth (magnetic activity).  Young ($\tau$ $<$ 300 Myr) G, K, 
and M stars are for the most part all fast rotators with rotation periods less 
than about 5 days.

\section{X-RAY OBSERVATIONS OF $\alpha$ CENTAURI B}

The $\alpha$ Centauri system was observed by the {\it R\"ontgen Satellite} 
({\it ROSAT}) High Resolution Imager (HRI; 0.1 - 2.4 keV), the {\it Chandra 
X-ray Observatory} ({\it Chandra}) Low Energy Transmission Grating 
Spectrometer (LETGS; 0.07 - 10 keV with the High Resolution Cameras, HRC-S and 
HRC-I), and the {\it XMM-Newton} ({\it XMM}) European Photon Imaging Camera 
(EPIC; 0.15 - 15 keV).  Since the range in measured coronal X-ray luminosity 
depends very strongly on the observed (or chosen) energy band, these 
luminosity measures cannot be directly compared.  Though $\alpha$ Cen B is a 
``soft'' coronal X-ray source, with no significant amounts of flux expected 
above about 1 keV, the overall energy passband used/assumed in determining 
X-ray luminosities is critical.

\begin{figure}
\epsscale{0.8}
\plotone{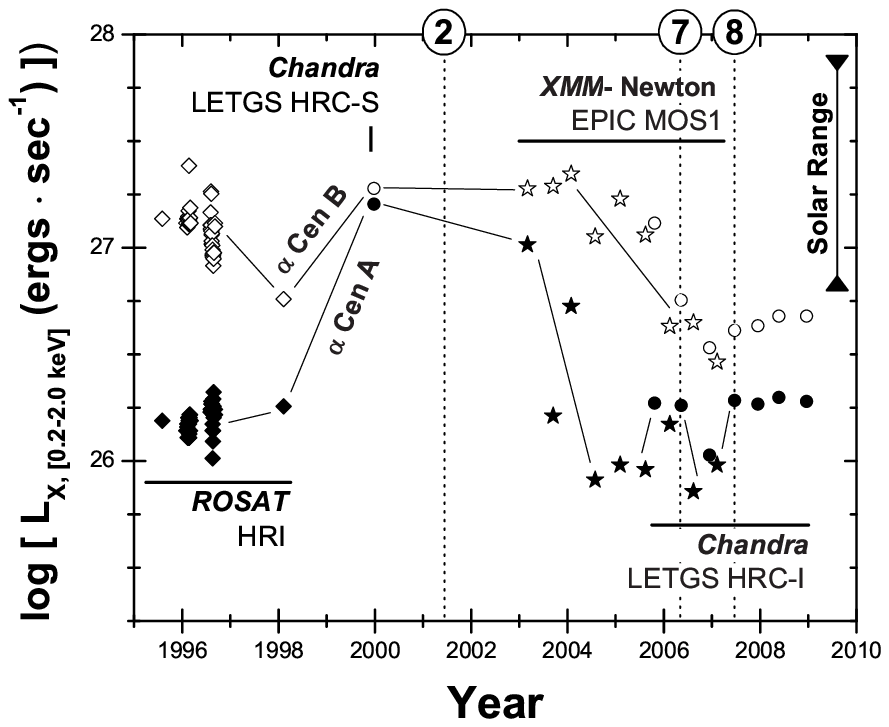}
\caption{Long-term coronal (soft) X-ray light curve of $\alpha$ Cen A (filled 
symbols) and $\alpha$ Cen B (open symbols).  The older {\it ROSAT} (1995-8; 
diamonds) and {\it Chandra} (1999 December; \protect\citealt{RNMMBK03}; circle) 
observations are combined with the more contemporary {\it XMM-Newton} 
(2003-2007; \protect\citealt{RSF05,RSH07}; stars) and {\it Chandra} (2005-2008; 
\protect\citealt{A09}; circles) observations.  Individual instrumental count rates 
were converted into the homogeneous (0.2 - 2.0 keV) passband using the 
appropriate energy conversion factors provided by \protect\citet{A09}.  Typical 
uncertainties for $\alpha$ Cen B ($\Delta$log{\it L}$^{\rm \,B}_{\rm \,X}$) 
are $\pm$0.017, $\pm$0.010, and $\pm$0.006, for the {\it ROSAT}, {\it XMM}, 
and {\it Chandra} measurements, respectively \protect\citep{A09}.  The vertical dotted 
lines with indicators along the top show when the {\it FUSE} Cycles 2, 7, and 
8 observations were obtained.  Note that the Cycle 7/8 observations coincide 
closely with a possible low magnetic activity state of $\alpha$ Cen B.  The 
vertical range bar on the right demonstrates the typical variability seen in 
the contemporary solar X-ray cycle \protect\citep{JSA03}, and discussed more 
completely in the text.  Noteworthy is that coronal X-ray maximum of the 
solar-like G2 V $\alpha$ Cen A roughly coincides with the minimum level of 
X-ray activity inferred for the current Sun. \label{fig:XRay}}
\end{figure}

Fortunately, an important recent study on the $\alpha$ Cen AB system (along 
with $\alpha$ CMi and $\epsilon$ Eri) by \citet{A09} nicely resolves this 
difficulty.  In this paper, all spatially resolved measurements from 
{\it ROSAT}, {\it Chandra}, and {\it XMM-Newton} are compared and studied on a 
common, homogeneous basis, and the appropriate conversion factors for the 
individual instrumental count rates to absolute fluxes are determined.  In our 
study, all of the original X-ray count rates have been converted into the 
homogeneous [0.2 - 2.0 keV] passband using the appropriate energy conversion 
factors provided by \citet{A09}.  Figure \ref{fig:XRay} displays these 
combined coronal X-ray luminosities ({\it L}$_{\rm \,X}$) for both $\alpha$ 
Cen A and $\alpha$ Cen B, respectively.  As reported by \citet{A09}, typical 
uncertainties ($\Delta${\it L}$_{\rm \,X}$) are approximately $\pm$4\%, 
$\pm$3\%, and $\pm$2\% for the individual {\it ROSAT}, {\it Chandra}, and 
{\it XMM} measurements, respectively.

Twenty-seven {\it ROSAT} pointings were carried out by J\"urgen Schmitt in 
1996 August/September and show only small-scale variability in the X-ray 
luminosity levels for $\alpha$ Cen B, averaging {\it L}$_{\rm \,X}$ = 1.1 
($\pm$ 0.26) $\times$10$^{\,27}$ erg s$^{-1}$.  At the time of the lone 
{\it Chandra} HRC-S observation in 1999 \citep{RNMMBK03}, the X-ray luminosity 
of $\alpha$ Cen B had slightly increased to {\it L}$_{\rm \,X}$ = 
1.9$\times$10$^{\,27}$ erg s$^{-1}$.  More recently, the {\it XMM} 
observations reported by \citet{RSF05} demonstrate that the coronal X-ray 
luminosity for $\alpha$ Cen B  diminishes from {\it L}$_{\rm \,X}$ = 
2.2$\times$10$^{\,27}$ erg s$^{-1}$ (early 2004) to {\it L}$_{\rm \,X}$ = 
0.3$\times$10$^{\,27}$ erg s$^{-1}$ (early 2007) -- a factor of 
$\sim$7$\times$ in three years' time.  This behavior is directly corroborated 
by the largely contemporaneously obtained {\it Chandra} observations 
\citep{A09}.  As seen in Figure \ref{fig:XRay}, $\alpha$ Cen B appears near 
the minimum of a coronal magnetic activity cycle around the early part of 
2007.

The range in solar X-ray luminosity also depends strongly on the observed (or 
chosen) energy band.  For example, existing estimations of the long-term 
variability of the Sun during its $\sim$11 year magnetic activity cycle are 
remarkably varied.  The Sun is also a ``soft'' coronal X-ray source, but the 
overall energy passband used/assumed in determining solar X-ray luminosities 
remains critical.  Using solar measurements from the {\it Student Nitric Oxide 
Explorer} ({\it SNOE}) satellite, \citet{JSA03} cite an expected true 
max-to-min variation of order 5-6$\times$ for the Sun over its full activity 
range for a typical contemporary cycle.  This stems from their analysis, and 
ultimate removal, of the effects of rotational modulation of active regions on 
the observed X-ray luminosities.  This reduces the total measured X-ray 
luminosity amplitude (12$\times$; activity cycle + rotational modulation) to 
that component that represents the changes due to the activity cycle alone 
(5-6$\times$).  In their study, great effort was made to convert the observed 
bandpass of the {\it SNOE} solar X-ray photometer (SXP; 0.04 - 0.6 keV) to 
that of the {\it ROSAT} Position Sensitive Proportional Counter (PSPC) 
bandpass (0.1 - 2.4 keV; in ``{\it ROSAT} All-Sky Survey'' mode).

On the other hand, the {\it Yohkoh} soft X-Ray telescope (SXT; 0.25 - 4.0 keV) 
also provides direct solar observations corresponding to those we consider 
here for $\alpha$ Cen B.  In his study, \citet{A96} determined that the Sun's 
activity between 1991 November (near solar maximum) and 1995 September (near 
solar minimum) exhibited a decline in the full-disk X-rays of greater than a 
factor of $\sim$25$\times$, with an accompanying change in average coronal 
temperature of only 3.3 to 1.5 MK.  However, it appears that X-ray energy 
sensitivity is strongly biased to high X-ray bandpasses, even though no 
significant amounts of solar flux are expected above about 1 keV.

An excellent review of previous and current published estimates of the Sun's 
X-ray luminosity is contained in the work of \citet{JSA03}.  Our Table 
\ref{tab:XSun} reproduces much of their effort here.  As seen, there is a 
large range of estimated total variation in the Sun's coronal X-ray luminosity 
over its long-term magnetic activity cycle, from about 4 - 50$\times$ 
(max/min)!  Further, these published ranges do not appear to be correlated to 
upper/lower energy sensitivities or overall X-ray bandpass.  Herein we adopt 
the result given by \citet{JSA03}.  We use their determination of the 
observed X-ray luminosity of the Sun, converted into the {\it ROSAT} PSPC 
bandpass, and extrapolated over the entire activity range of a typical 
contemporary solar cycle.  This overall variability (12$\times$) retains the 
additional influence of rotational modulation.  Since it is not possible to 
remove all effects of rotational modulation on the observed X-ray luminosities 
of $\alpha$ Cen A and B due to incomplete temporal coverage (Figure 
\ref{fig:XRay}), in this manner a direct comparison with the Sun can still be 
made.

\begin{figure}
\vspace{3cm}
\plotone{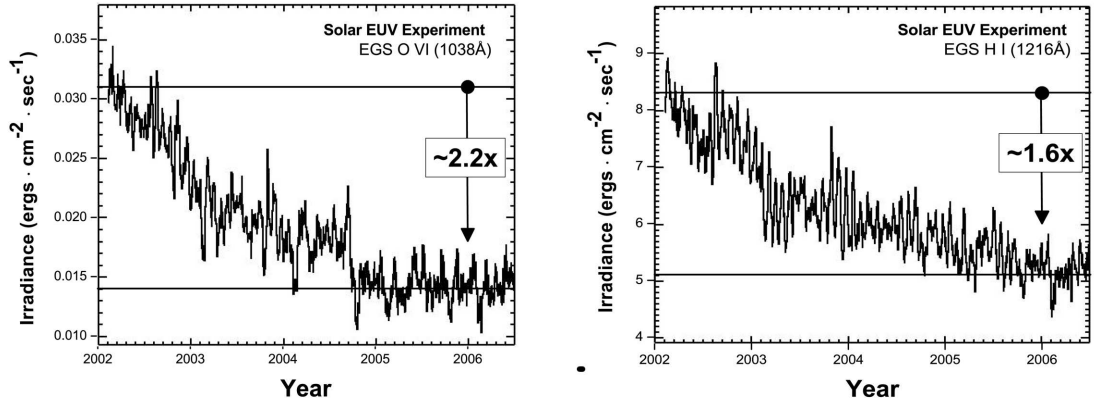}
\caption{Left panel:  {\it SEE} EGS observations of TR-originating 
\ion{O}{6} emission fluxes for the Sun show a decline of $\sim$2.2$\times$ 
from near solar max (2002) to solar min.  Right panel:  the Sun declines 
by a factor of $\sim$1.6$\times$ in the chromospherically produced \ion{H}{1} 
Ly$\alpha$ emission levels over its $\sim$11 year magnetic activity cycle.  
Both of these declines are primarily due to variations in the overall level of 
magnetic activity.  The Sun is also observed to vary by a factor of about 
$\sim$12$\times$ in X-ray luminosity (see the text) over the same timeframe, 
thereby exhibiting a high degree of correlation in emission measures throughout 
its entire atmospheric structure. \label{fig:SEE}}
\end{figure}

\citet{JSA03} demonstrate that our Sun typically fluctuates in coronal X-ray 
luminosity by a factor of $\sim$12$\times$ ({\it L}$^{\rm ~Sun}_{\rm \,X}$ 
$\approx$ 6.3-79 $\times$10$^{\,26}$ erg s$^{-1}$) over its $\sim$11 year 
magnetic activity cycle.  Since this X-ray variability is primarily due to 
changes in the level of overall magnetic activity, not merely coronal 
temperature changes, the corresponding TR and chromospheric variations are 
both expected and measured to be tightly correlated (see \citealt{L97}).  
Indeed, {\it Solar EUV Experiment} ({\it SEE}, aboard NASA's {\it TIMED} 
mission; see \citealt{WEBCLRSTW05}) observations from 2002 (near solar 
maximum) to 2006 (solar minimum) show a decline in the TR-produced \ion{O}{6} 
($\lambda$1038\AA) emission feature of $\sim$2.2$\times$, and a decline of 
$\sim$1.6$\times$ in the chromospheric \ion{H}{1} Ly-$\alpha$ emission levels 
(Figure \ref{fig:SEE}).

Studies of cosmogenic isotopes, C$^{14}$ (in tree rings) and Be$^{10}$ (in ice 
cores), have been employed to trace the Sun's activity back over the last 
$\sim$12,000 years \citep{SUKSB04}.  Curiously, this study reveals that the 
solar magnetic activity (defined by solar winds and sunspots) appears to have 
been exceptionally high over the last $\sim$70 years, but is perhaps declining 
now.  This may help to explain the apparently low (relative to the present 
Sun) coronal soft (0.1 - 2.5 keV) X-ray luminosities observed for the solar 
analog $\alpha$ Cen A, as seen in Figure \ref{fig:XRay}.  Three other 
well-studied solar-type stars (18 Sco, 16 Cyg A and B) also appear to have 
X-ray emissions considerably lower than current solar values.  18 Sco is the 
best known match to the Sun with regard to age, rotation, and physical 
properties, and has an observed {\it L}$^{\rm ~18Sco}_{\rm \,X}$ $\approx$ 
8 $\times$10$^{\,26}$ erg s$^{-1}$ at the maximum of its 7-11 year activity 
cycle \citep{CGEDHDT10}.  This is comparable to the maximum value of 
{\it L}$_{\rm \,X}$ for $\alpha$ Cen A, but less than one half of the modern 
(last two decades) solar values.  We have currently been granted {\it Chandra} 
X-ray observations of 18 Sco during its activity cycle minimum.  Again, as in 
the case of $\alpha$ Cen A, this lends support to the hypothesis that the Sun 
is in an exceptional high activity level compared to historical levels, defined 
over the last 10,000 years.  Although too soon to know for sure, based on 
previous and current solar activity cycles/measures, the Sun may show signs of 
declining activity (see \citealt{KDSDK10}; NOAA Space Weather Prediction 
Center; \url{www.swpc.noaa.gov/SolarCycle/}).

It is clear that a direct comparison of the Sun to $\alpha$ Cen B is 
problematic, since different instruments (and X-ray bandpasses) are employed 
for the two disparate objects (see, for example, \citealt{MM03}).  This 
notwithstanding, the X-ray [0.2 - 2.0 keV] luminosity of $\alpha$ Cen B has 
decreased by a factor of nearly 7$\times$ over the $\sim$5 year time frame of 
the self-consistent {\it XMM} and/or {\it Chandra} observations.  This 
behavior thus appears consistent with a solar-like $\sim$11 year X-ray 
activity cycle.

\subsection{Long-term Coronal X-ray Activity Cycle}

The entire $\alpha$ Cen B X-ray dataset spans $\sim$13 years (1995-2008) 
making it potentially sufficient, temporally, to determine the long-term 
coronal magnetic activity cycle.  Therefore, all available X-ray luminosity 
observations were analyzed using the method of Lomb-Scargle \citep{L76,S82}, 
as modified for unevenly sampled data (see \citealt{HB86,PR89}).  This 
particular algorithm has certain computational advantages over other methods, 
such as the treatment of missing values and a quantitative estimate of the 
false discovery rate (FDR).  In Section 6 below, additional discussion of the 
``robustness'' of these analytically-determined FDR's is given.  The results 
were further refined with an iterative grid search procedure to determine 
possible cycle rates, modulation amplitudes, and phase information for 
$\alpha$ Cen B.

\begin{figure}
\epsscale{0.8}
\plotone{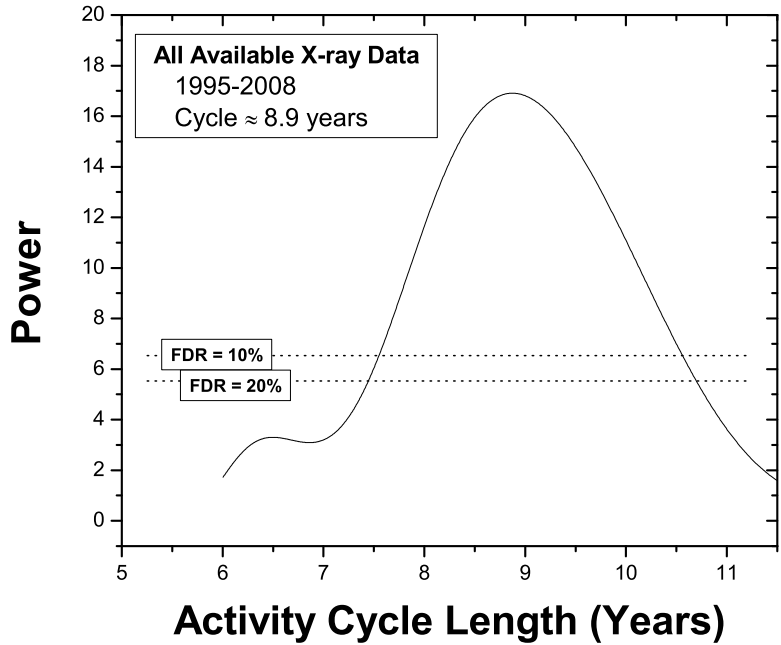}
\caption{Lomb-Scargle power spectrum for all available (1995-2008; {\it 
ROSAT}, {\it Chandra}, {\it XMM}) X-ray data for $\alpha$ Cen B.  All of the 
individual instrumental count rates were converted into the homogeneous 
(0.2 - 2.0 keV) passband using the appropriate energy conversion factors 
provided by \citet{A09}.  As seen, though cycle rates between 7.5 - 10.75 
years would be plausible with better than 20\% FDR, a long-term magnetic 
activity cycle of $\sim$8.9 years is favored with an extremely low FDR.  The 
broad nature of the power spectrum is due to the limited temporal range of 
data -- slightly longer than one complete cycle. \label{fig:LongX}}
\end{figure}

Figure \ref{fig:LongX} shows the resultant Lomb-Scargle power spectrum for 
this complete (1995-2008; {\it ROSAT}, {\it Chandra}, {\it XMM}) set of 
coronal X-ray measurements for $\alpha$ Cen B.  Unfortunately, with only 
slightly more than one complete cycle's worth of X-ray data, the long-term 
magnetic activity cycle length is not well constrained.  As seen from the 
relatively broad nature of the power spectrum, cycle rates of between about 
7.5 and 10.75 years must be considered plausible with better than 20\% FDR.  
This notwithstanding, a cycle length of {\it P}$_{\rm cycle}$ $\approx$ 8.9 
years is favored with an apparent very high confidence (FDR is essentially 
0\%!).

\subsection{Short-term Coronal X-ray Modulation - Rotation Period}

To determine the rotation period of $\alpha$ Cen B from the low-amplitude, 
sinusoidal-like modulation of light due to the rotational effects attributed 
to the presence of magnetically active regions (starspots, plages, bright 
light faculae, chromospheric or coronal features), appropriate sections of 
data need to be isolated.  Ideally, suitable portions would be $\sim$50 - 100 
days long -- long enough to contain sufficient observations, but short enough 
to minimize the effects of possible formation, migration, and/or destruction 
of these active regions.  Fortunately, the 1996 August-September {\it ROSAT} 
pointings contain 27 individual X-ray measures spanning $\sim$35 days that 
should be sufficient for the determination of the rotation period of $\alpha$ 
Cen B.

\begin{figure}
\epsscale{0.8}
\plotone{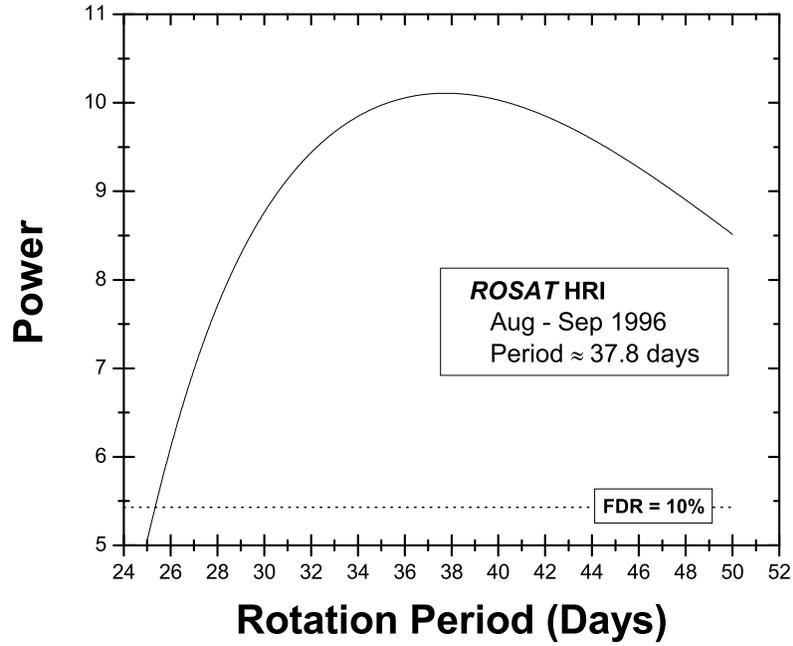}
\caption{Lomb-Scargle power spectrum for the 1996 August-September {\it ROSAT} 
HRI X-ray luminosity data for $\alpha$ Cen B.  Apparent in this particular 
section of data is the modulation of light due to the presence of magnetically 
active regions.  Though there is very high confidence in a $\sim$37.8 day 
rotation period, the broad nature of the power spectrum is due to the limited 
temporal range of information -- less than one complete rotation in length. 
\label{fig:ShortX}}
\end{figure}

\begin{figure}
\epsscale{0.8}
\plotone{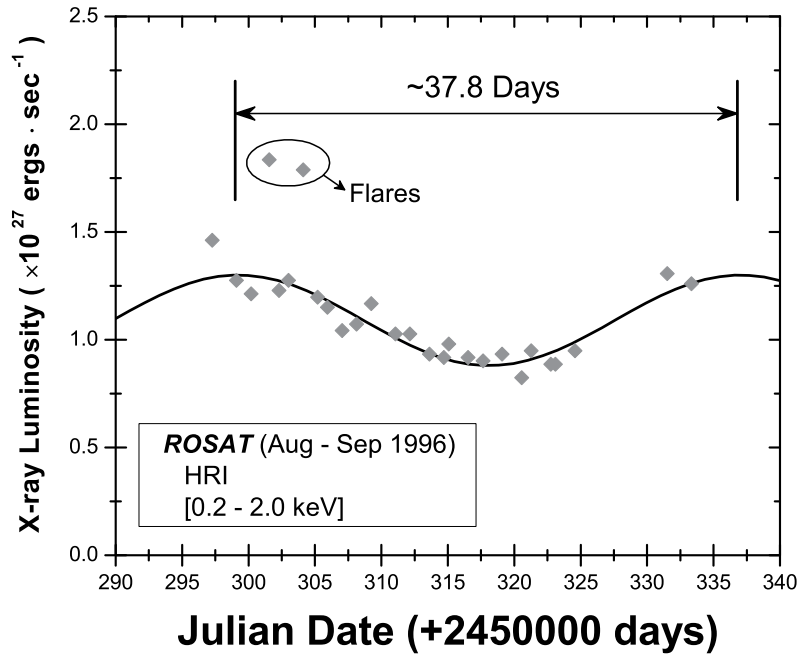}
\caption{1996 August-September {\it ROSAT} HRI observations (gray diamonds) 
shown above are overplotted with the appropriately modulated and phased light 
curve, as determined utilizing an iterative grid search method and 
incorporating the $\sim$37.8 day rotation period ascertained from the above 
(see Figure \ref{fig:ShortX}) Lomb-Scargle analysis. \label{fig:ShortXRay}}
\end{figure}

Figure \ref{fig:ShortX} shows the resultant Lomb-Scargle power spectrum for 
this subsection of {\it ROSAT} X-ray observations for $\alpha$ Cen B.  As 
seen, a relatively large range (25.3 - 50+ days) would result in an estimated 
FDR better than 10\%, but a rotation period of {\it P}$_{\rm rotation}$ 
$\doteq$ 37.8 days is favored with an apparent very high confidence (FDR 
$\approx$ 0.1\%).  Figure \ref{fig:ShortXRay} shows the 1996 August-September 
X-ray luminosity observations with the best-fitting, 37.8 day period, light 
curve overplotted.  An iterative grid search routine was employed to determine 
the appropriate light modulation amplitude and cycle phase.  In this case, 
with less than one complete cycle's worth of information, this rotation period 
determination cannot be considered well constrained, but in fact is consistent 
with other independent period estimates (35 - 42 days) as addressed in more 
detail below (Section 5.2).  As in the case of the Sun, differential rotation 
of active regions could produce year-to-year variations in observed rotation.

\section{{\it FUSE} FAR-ULTRAVIOLET OBSERVATIONS OF $\alpha$ CENTAURI B}

The key emission line fluxes obtained by the {\it Far Ultraviolet 
Spectroscopic Explorer} ({\it FUSE}) satellite are used to probe specific 
regions of the stellar atmosphere, from the hot plasmas of the upper 
chromosphere (\ion{H}{1} Lyman $\alpha\beta\gamma$... series, $\sim$12,000 K; 
\ion{C}{2} ($\lambda$1036/7\AA), $\sim$20,000 K), the TR (\ion{C}{2}; 
\ion{C}{3} ($\lambda$977,$\lambda$1176\AA), $\sim$50,000 K; \ion{O}{6} 
($\lambda$1032,$\lambda$1038\AA), $\sim$300,000 K) through the low corona 
(\ion{O}{6}).

The {\it FUSE} satellite observed both $\alpha$ Cen A and $\alpha$ Cen B on 
three separate occasions (see Table \ref{tab:FUSELog}).  In 2001 June (Cycle 
2), both stars were observed with the medium resolution aperture (MDRS, 
4$\times$20$\arcsec$; Program ID: P104; PI: K. Sembach).  The spectra have 
excellent signal-to-noise ratios (S/Ns) for the stronger emission features, 
and the peak flux of the \ion{C}{3} ($\lambda$977\AA) line for $\alpha$ Cen B 
was $\mathscr{F}_{\rm \,C III,{\lambda}977}$ = 4.1$\times$10$^{-11}$ (cgs).  
In 2006 May (Cycle 7) and 2007 June (Cycle 8), $\alpha$ Cen A and B were again 
observed (ID: G081 and H096; PI: L. DeWarf).  At these times, the integrated 
emission flux values of the same \ion{C}{3} emission feature for $\alpha$ 
Cen B had diminished to 1.7$\times$10$^{-11}$ and 2.1$\times$10$^{-11}$ (cgs), 
respectively -- dropping by a factor of $\sim$2.5$\times$ in about five years' 
time.

The angular separation between $\alpha$ Cen A and B during the Cycle 7 (2006 
May 5) observations had closed to 9$\,\farcs$53 along a position angle of 
$\theta$ = 233$^{\circ}$.  Since the failure of two reaction wheels on the 
{\it FUSE} spacecraft in 2001 December, this provided some additional 
difficulties for pointing and acquisition.  To obtain separate spectra for 
$\alpha$ Cen A and B using the MDRS aperture, the roll angle of the spacecraft 
needed to be very nearly perpendicular to the position angle of the two stars 
on the sky -- a maneuver rendered significantly more problematic for the 
crippled satellite.  Fortunately, through great effort by the {\it FUSE} team, 
the individual spectra of $\alpha$ Cen A and B were ultimately secured.

All of the $\alpha$ Cen A and B data were processed in a uniform manner, 
utilizing the most recent CalFuse calibration pipeline (version 3.2.0).  
For consistency, the pre-existing 2001 data have also been reprocessed with 
this same CalFuse version.  Unfortunately, certain individual exposures 
obtained no discernable flux from $\alpha$ Cen B.  Another ``idiosyncrasy'' 
in the individual exposures were partial flux levels.  These anomalously low 
flux exposures were most likely the result of the star being intermittently 
contained within the aperture during the duration of the exposure.  When 
examined individually, we found a few of these flawed exposures in the 
pre-existing 2001 data as well.  Therefore, an exposure-by-exposure, 
segment-by-segment analysis of all of the $\alpha$ Cen B {\it FUSE} data was 
necessitated and all faulty exposures were removed.  Fortunately, due to the 
proximity of the $\alpha$ Cen system resulting in high received FUV flux 
values, the S/Ns of the remaining individual ``On Target'' exposures were 
excellent.

Table \ref{tab:FUVFlux} provides the integrated flux values measured for the 
key FUV emission features for all three of the {\it FUSE} observing cycles.  
The uncertainty estimates, stemming from the internal inconsistencies among 
the individual segments and exposures for a given feature, are also listed.  
We find that, on average, the individual {\it FUSE} spectra, secured during a 
given epoch, are consistent to better than about $\pm$25\%.

\begin{figure}
\epsscale{0.7}
\plotone{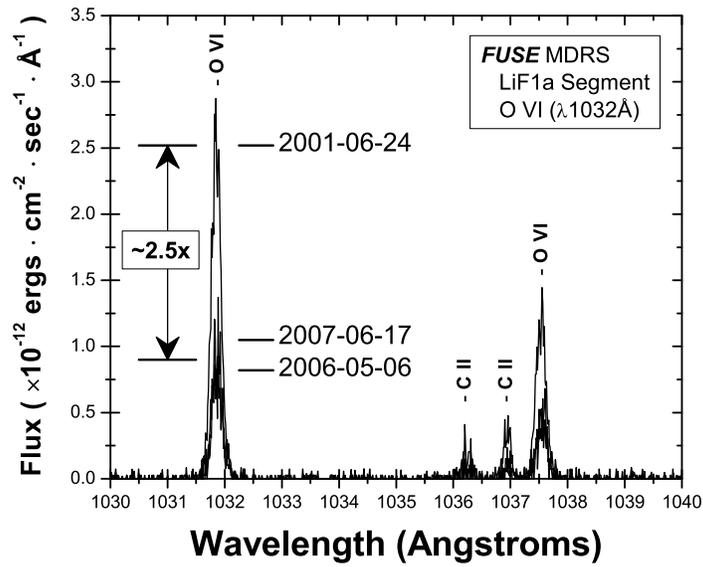}
\caption{Comparison of representative {\it FUSE} MDRS LiF1a spectra obtained 
in 2001 June, 2006 May, and 2007 June for $\alpha$ Cen B.  Note the excellent 
S/N of these individual exposures, though as discussed in the text, the 
individual {\it FUSE} spectra, secured during a given epoch, are only 
internally consistent to about $\pm$25\%.  Clearly seen is the significant 
drop by a factor of $\sim$2.5$\times$ in the integrated emission fluxes for 
the TR/chromospheric transitions (\ion{O}{6}, \ion{C}{2}) over the five year 
span from 2001 to 2006.  The substantial decline of these key FUV emission 
measures is likely indicative of a deep-rooted magnetic activity change. 
\label{fig:CompareB}}
\end{figure}

Figure \ref{fig:CompareB} shows the comparison between the emission levels for 
$\alpha$ Cen B for representative {\it FUSE} exposures at the three different 
epochs.  Note the substantial decline in integrated FUV flux levels from 2001 
June to 2006 May.  The ratio of change in these integrated emission flux 
values, with respect to the 2001 (Cycle 2) values are listed in Table 
\ref{tab:FluxRatio}.  As seen, during the span of time between the Cycle 2 
and the Cycle 7/8 observations, $\alpha$ Cen B diminishes by a factor 
of $\sim$2.3-3.3$\times$ in all of these key TR/chromospheric emission 
measures.  Further, the higher temperature transitions (\ion{O}{6}) diminish 
by a slightly larger amount than the lower temperature transitions 
(\ion{C}{2}, \ion{C}{3}).  A similar trend is observed during the solar 
activity cycle (see \citealt{L91,L97,A96}), and it is expected that as 
$\alpha$ Cen B becomes less active, changes in starspots, faculae, plages, and 
magnetic network modify the net radiative output by altering the temperature 
and density of the otherwise homogeneous stellar atmosphere.  Unfortunately, 
given our estimates for the consistency within the individual {\it FUSE} 
spectra, little more can be said about this likely atmospheric temperature 
trend.  The overall drop in FUV emission levels, though, is very likely 
indicative of a deep-rooted long-term magnetic activity cycle, since 
variations in atmospheric temperature alone would result in greater departures 
in the relative differences for the respective atomic species.  That is, the 
reasonably large change in temperature that would be required to alone produce 
the factor of about 7$\times$ difference in the coronally-produced X-ray 
luminosities presented in Section 3, should also result in a more substantial 
change in certain FUV temperature sensitive ratios, such as 
($\mathscr{F}_{\rm \,O\,VI}$ / $\mathscr{F}_{\rm \,C\,II}$).  On the other 
hand, a decline in the overall level of magnetic activity in the star would be 
expected to reduce commensurately all emission fluxes, as has been addressed 
above for the case of the Sun.

\begin{figure}
\epsscale{0.7}
\plotone{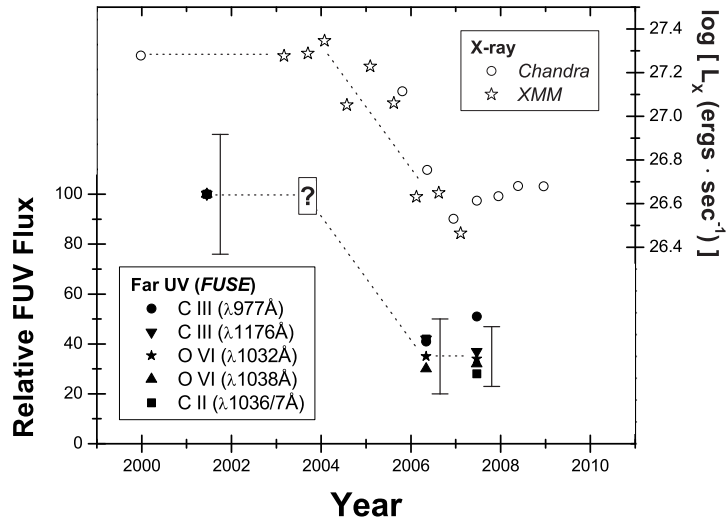}
\caption{Shown in the lower panel above is the relative change in the key FUV 
emission line fluxes over the three epochs observed by {\it FUSE} for $\alpha$ 
Cen B.  All FUV measures are normalized to the 2001 (Cycle 2) observations and 
the mean individual uncertainty ranges are shown slightly to the right of each 
epoch.  Note the significant drop ($\sim$2.3-3.3$\times$) over the five year 
span from 2001 to 2006 for all of these TR/chromospheric emission features.  
As expected, there appears to be a slight temperature trend -- the higher 
temperature transitions (\ion{O}{6}) diminish by a potentially larger amount 
than the lower temperature transitions (\ion{C}{2}, \ion{C}{3}).  That is, as 
$\alpha$ Cen B becomes less active its mean atmospheric temperature becomes 
somewhat cooler.  These observed changes in the diagnostic FUV emission line 
fluxes appear to be in direct accord with the changes taking place in the 
corona (X-ray; upper panel), pointing to a deep-rooted change in the level of 
magnetic activity over the possible $\sim$8.9 year cycle of $\alpha$ Cen B. 
\label{fig:FUSEFlux}}
\end{figure}

In Figure \ref{fig:FUSEFlux}, the relative fluxes of the key FUV emission 
lines, as determined with {\it FUSE}, are displayed over the same time period 
as the above X-ray observations.  This corresponding drop in the FUV 
emissions, throughout the entire atmospheric structure of the star, indicate a 
true decrease in the level of magnetic activity, not merely a coronal 
temperature change.

\subsection{Electron Pressures in the Transition Region}

We carried out measurements of TR plasma electron pressures ({\it P}$_e$) 
using the same method described by \citet{GRH03} in a {\it FUSE} study of 
solar analogs.  One interesting preliminary result for $\alpha$ Cen B, derived 
from our analysis of the TR electron density sensitive ${\cal R}$ $\equiv$ 
$\mathscr{F}_{\rm \,C III,{\lambda}1176}$ / $\mathscr{F}_{\rm \,C III,{\lambda}977}$ 
emission line ratio, infers that the {\it P}$_e$ appears to remain essentially 
unchanged during the decline in magnetic activity defined by the X-ray fluxes 
from 2001 to 2006, but then drops significantly from 2006 to 2007, during 
magnetic quiescence.  We find that at the magnetic activity ``high-state'' of 
2001, ${\cal R}$ $\approx$ 0.28, resulting in a log$P_e$ $\approx$ 14.15 
(cgs).  At the beginning of the ``low-state'' in 2006, ${\cal R}$ $\approx$ 
0.29, but then appears to abruptly decrease to ${\cal R}$ $\approx$ 0.16 by 
2007, resulting in an inferred lower electron pressure, log{\it P}$_e$ 
$\approx$ 13.5 (cgs).

By comparison, the ``active'' Sun (by measuring discrete active regions of the 
Sun, such as sunspots/plage regions) has an estimated \ion{C}{3} 
${\cal R}$-value of about 0.44 \citep{NRDK85,DRNK85}, with the ``quiet'' Sun 
at 0.29 \citep{DFJ76}.  Curiously, the proportional change for both $\alpha$ 
Cen B and the Sun are comparable, i.e., 
(${\cal R}_{\rm active}$/${\cal R}_{\rm quiescent}$) $\approx$ 1.67.  It 
should be kept in mind, though, that these {\it FUSE} measures of ${\cal R}$ 
are integrated over the star's visible disk and would therefore be influenced 
by the distribution and the number of active emitting regions of differing 
sizes and activity strengths.

It would have been interesting to confirm this result for $\alpha$ Cen B and 
follow possible {\it P}$_e$ changes over more of this stars' activity cycle, 
but with the demise of the {\it FUSE} satellite, this will no longer be 
possible in the FUV bandpass.

\section{ULTRAVIOLET OBSERVATIONS OF $\alpha$ CENTAURI B}

The {\it International Ultraviolet Explorer} ({\it IUE}) satellite covered 
wavelengths from 1150 to 3200\AA, allowing us to analyze changes in the 
prominent \ion{C}{4} ($\lambda$1550\AA) and \ion{Mg}{2} {\it h+k} 
($\sim$$\lambda$2800\AA) emission features from $\alpha$ Cen B.  The 
\ion{C}{4} emission occurs at temperatures from 50,000 to 100,000 K, typically 
found in the TR of the star.  \ion{Mg}{2} (8000-12,000K) typically originates 
in the stellar chromosphere.  The longevity of the satellite results in a very 
large dataset of UV spectral data, critical to long temporal studies.  We have 
utilized observations from 1978 August to 1995 July, providing us with an 
extensive timeline of \ion{C}{4} and \ion{Mg}{2} information.

\begin{figure}
\epsscale{0.8}
\plotone{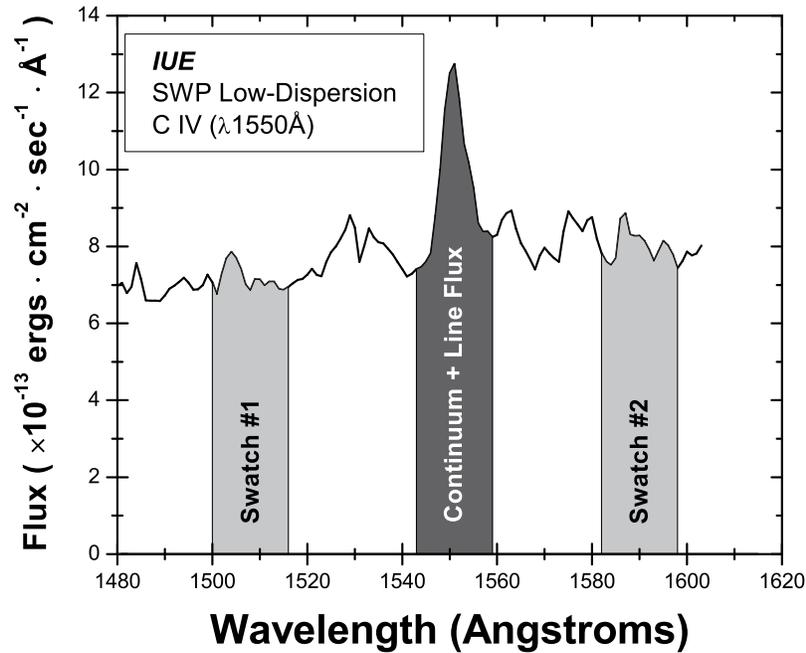}
\caption{Shown above is a typical {\it IUE} SWP low-dispersion spectra of 
$\alpha$ Cen B that diagrams the swatches (as discussed more completely in 
the text) used to determine the integrated line and continuum fluxes for 
the \ion{C}{4} ($\lambda$1550\AA; 50,000-100,000K; TR) emission feature. 
\label{fig:IUECIV}}
\end{figure}

For $\alpha$ Cen B, there are $\sim$90 low-dispersion ($\sim$6-7\AA), large 
aperture (9$\times$22$\arcsec$) {\it IUE} spectra obtained by the satellite's 
short wave primary (SWP; 1150-1980\AA) camera.  Focusing on changes in the 
TR-produced \ion{C}{4} ($\lambda$1550\AA) emission line over time, the 
spectra were reduced by first quadratically interpolating the intensity to 
each integer wavelength.  Next, the center of the \ion{C}{4} line core was 
ascertained by searching for the maximum intensity value between the 
wavelength range of 1530 and 1565\AA.  The flux was then summed over a 16\AA\ 
swatch around this peak value ($\pm$8\AA), resulting in integrated flux values 
for the \ion{C}{4} line, combined with the contribution of the continuum in 
this bandpass.  To account for, and ultimately remove, this continuum flux 
from the \ion{C}{4} data, an average of the total flux of two bracketing 
16\AA\ swatches (1500-1516\AA\ and 1582-1598\AA) was then subtracted from the 
above line+continuum flux.  This provided us with an accurate determination of 
the integrated \ion{C}{4} line emission for $\alpha$ Cen B.  Figure 
\ref{fig:IUECIV} diagrams both the \ion{C}{4} feature and the swatches used 
to determine the integrated line and continuum fluxes with a typical {\it IUE} 
SWP low-dispersion, large-aperture spectra.  Table \ref{tab:IUESWP} lists the 
complete results.

\begin{figure}
\epsscale{0.8}
\plotone{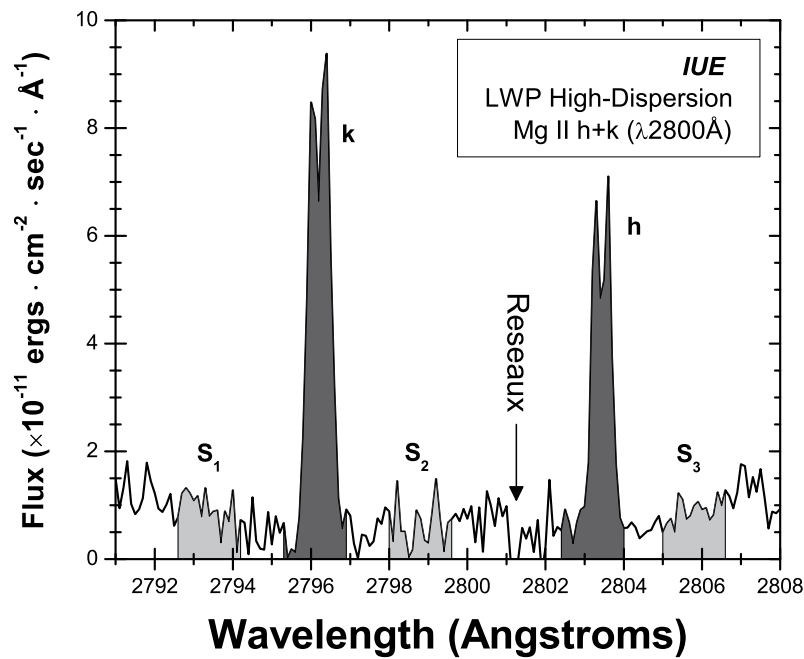}
\caption{``Swatch'' method similar to that used for the \ion{C}{4} analysis 
({\it see} Figure \ref{fig:IUECIV}) was implemented to determine emission flux 
strengths for the \ion{Mg}{2} {\it h+k} ($\lambda$2803+2796\AA; 8,000-12,000K; 
chromosphere) emission features.  As described more thoroughly in the text, we 
have defined an \ion{Mg}{2} Index by dividing the combined integrated fluxes 
found in the {\it h} and {\it k} emission line cores by a weighted average 
summation of integrated continuum values (swatches). \label{fig:IUEMgII}}
\end{figure}

To analyze temporal variations in the chromospherically produced \ion{Mg}{2} 
{\it h+k} ($\lambda$2803+2796\AA) emission features over the same timeline, 
this study also included $\sim$60 high resolution (0.1-0.3\AA), large aperture 
(10$\times$23$\arcsec$) spectra from the {\it IUE} satellite's long wave 
primary (LWP) and long wave redundant (LWR) cameras.  Due to differences in 
the profile of these spectral features, it was necessary to reduce these data 
in a different manner.  We used fixed wavelength regions to define the 
individual \ion{Mg}{2} {\it h+k} line cores and three separate swatch 
regions.  An \ion{Mg}{2} ``Index'' was defined by dividing the integrated flux 
values for each line by a combination of the three continuum swatches, i.e.
\begin{eqnarray}
{\rm MgII\,h+k\,emission\,index}\,\,=\,\,\frac{h+k}{S}\,, \nonumber \\ 
\,\,{\rm where}\,S\,=\,\left( \frac{S_1 + 2S_2 + S_3}{4} \right)\,. \nonumber
\end{eqnarray}
\noindent In the above relation, the values {\it h} and {\it k} refer to the 
total integrated flux in the region of the emission features, 
2802.4-2804.0\AA\ and 2795.3-2796.9\AA, respectively.  The three swatches are 
defined to be {\it S}$_{1}$ (2792.6-2794.2\AA), {\it S}$_{2}$ 
(2798.0-2799.6\AA), and {\it S}$_{3}$ (2805.0-2806.6\AA).  The swatch 
{\it S}$_{2}$ is not centered directly between the {\it h+k} features to avoid 
the fiducial mark (reseaux) incorporated into the {\it IUE} vidicon detector.  
An example of this method for measuring the \ion{Mg}{2} {\it h+k} emission 
line strength is shown in Figure \ref{fig:IUEMgII}, and Table \ref{tab:IUELWP} 
lists the complete results.

The {\it IUE} vidicon cameras had a maximum dynamic range of 256, though this 
was significantly limited by background noise and a low-level baseline 
intensity ``pedestal.''  \citet{A93} point out that the maximum intrinsic S/N 
per pixel is $\sim$12:1.  Due to the manner by which adjacent pixels were 
grouped in low resolution mode, \citet{MF00} demonstrate that an S/N 
$\lesssim$ 30:1 is ideally possible per spectral element.  They also point to 
substantial systematic effects that remain in the {\it IUE} New Spectroscopic 
Image Processing System (NEWSIPS) data (used in this study), producing spectra 
that are only internally consistent to 10\%-15\%, at best.  Unfortunately, the 
majority of these remaining systematics are time dependent.  The flux 
sensitivity of the LWR camera, the worst case example, is shown to diminish by 
about 10\% over its $\sim$5 years of service.  Short-term temporal systematics 
also remain in the data, and can contribute to a ``noise-like'' uncertainty 
exceeding 10\%.

In an attempt to assess the overall quality of the {\it IUE} data presented 
here, as it was reduced using the techniques outlined above, we selected 
several UV flux standard stars that were used as calibration standards for 
{\it HST} (see \citealt{BHHG90}).  We measured all available spectra for 
the two stars that were the closest spectral type matches to $\alpha$ Cen B, 
HD 186427 (G5 V) and HD 27836 (G1 V), along with the cool subgiant HD 2151 
(G1 IV).  We also subjected our reduction routines to sample spectra of the 
two {\it IUE} ``prime standard'' stars, HD 50753 (B3 IV) and HD 93521 (O9 V).  
With regards especially to the latter prime standard stars, these spectra were 
obtained uniformly throughout the life of the satellite, and therefore address 
the extent of the potentially remaining long- and short-term systematics in 
the NEWSIPS data, along with the internal consistencies (noisiness) within the 
individual {\it IUE} LWP/R and SWP spectra.

From this analysis we find that the internal consistency of the {\it IUE} 
NEWSIPS data, convolved with our method of data reduction, results in an 
uncertainty of better than about 20\% for our integrated \ion{C}{4} line 
emission measures, and  $\lesssim$10\% for the individual \ion{Mg}{2} 
{\it h+k} emission Index values.  In addition, there were no discernable 
trends in the reduced spectra that were common to all of the measured standard 
stars that could have given rise to potential ``false-positive'' cycle 
determinations within the $\alpha$ Cen B data set.

\subsection{Long-term Magnetic Activity Cycle, Revisited}

\begin{figure}
\epsscale{0.8}
\plotone{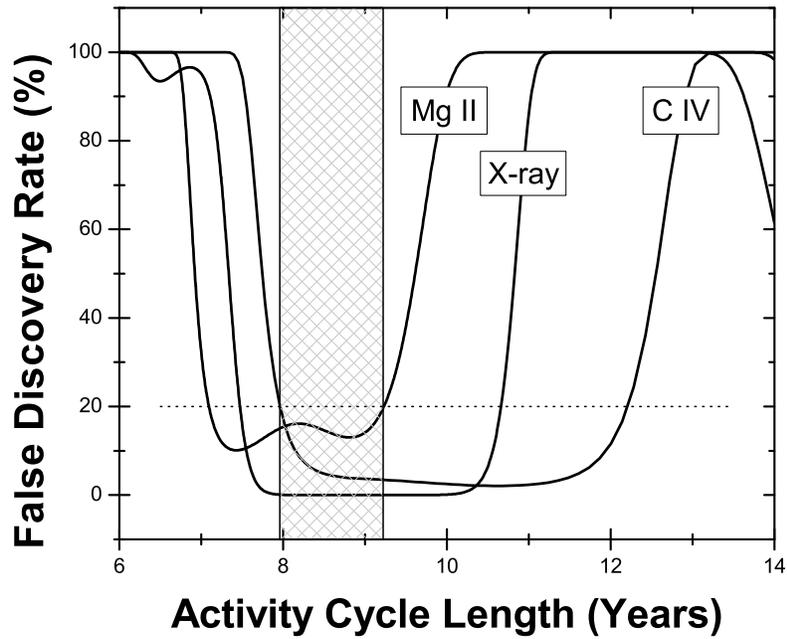}
\caption{Long-term activity cycle false discovery rates (FDR) for all 
of the available {\it IUE} \ion{C}{4} emission and \ion{Mg}{2} Index 
(1978-1995) data, along with the X-ray ({\it ROSAT}, {\it Chandra}, {\it XMM}) 
luminosity observations.  Due to the temporally spasmodic nature of the 
acquired $\alpha$ Cen B observations, narrow confinement of the overall 
activity cycle is not strictly possible, but considering the intersection of 
all FDR results together point to a possible range of about 7.96 - 9.22 years, 
with a better than 20\% FDR. \label{fig:LongFDR}}
\end{figure}

The long-lived (1978-1995) {\it IUE} satellite provides an opportunity to 
study $\alpha$ Cen B's magnetic activity over a relatively long period in 
the years preceding the current magnetic quiescence, as defined by the X-ray 
and FUV data described above (Sections 3 and 4).  The results of the 
Lomb-Scargle periodogram analysis, with the complete \ion{C}{4} 
($\lambda$1550\AA) and \ion{Mg}{2} {\it h+k} ($\sim$$\lambda$2800\AA) data 
sets along with the entire X-ray luminosity measures, are shown in Figure 
\ref{fig:LongFDR}.  If we require an FDR of better than 20\%, the intersection 
of the three FDR functions results in an expected long-term activity cycle 
between 7.96 and 9.22 years.

In their analysis of the \ion{Mg}{2} data, \citet{BM08} determine a long-term 
activity cycle of 8.38 yr (3061 days).  Their procedure involved transforming 
the mean continuum flux near the \ion{Mg}{2} {\it h+k} emission lines to the 
corresponding Mount Wilson \ion{Ca}{2} HK $<S>$ index, which was then 
combined with optically determined values.

\begin{figure}
\epsscale{0.8}
\plotone{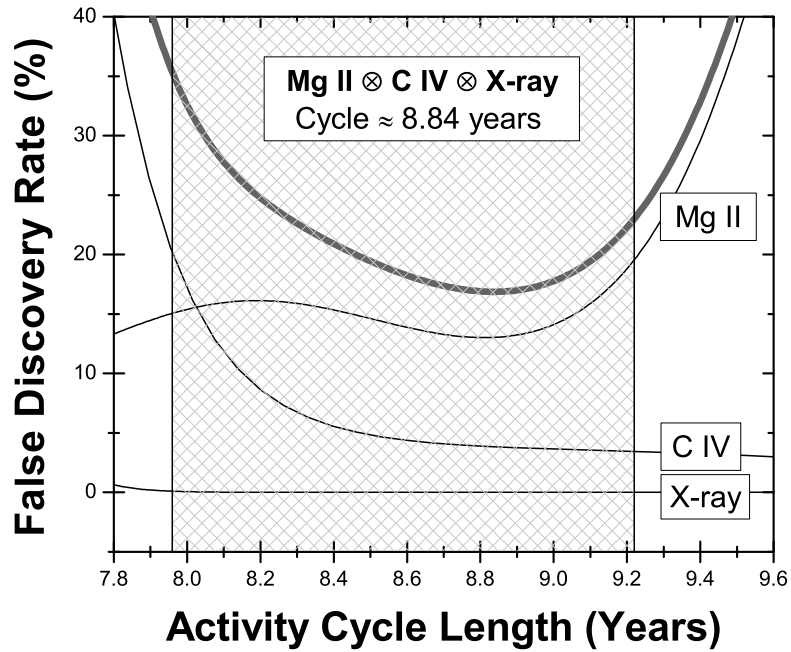}
\caption{Convolution of the \ion{C}{4} emission, \ion{Mg}{2} Index, and 
X-ray luminosity FDR functions result in a best determination of $\sim$8.84 
yr (3230 days) for the long-term magnetic activity cycle for $\alpha$ Cen B. 
\label{fig:LongConvolve}}
\end{figure}

In this study, the convolution of all FDR distributions (see Figure 
\ref{fig:LongConvolve}) results in an ``ultimate'' best determination of 
{\it P}$_{\rm cycle}$ $\approx$ 8.84 years (3230 days) for the long-term 
activity cycle for $\alpha$ Cen B.  It should be noted, though, that strict 
comparison of the FDR functions for the individual X-ray and {\it IUE} 
data sets can be somewhat challenging.  Though the {\it IUE} data is 
quantifiably noisier (producing less favorable FDRs), it encompasses multiple 
complete cycles (producing a much better constrained cycle duration).

\begin{figure}
\epsscale{0.7}
\plotone{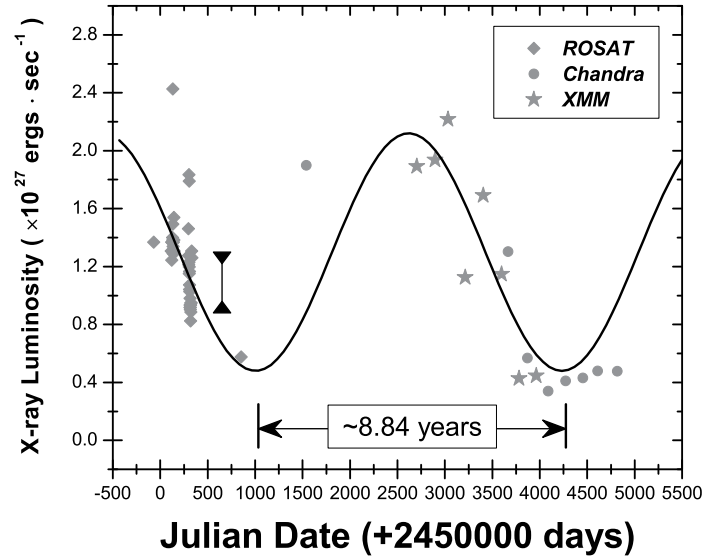}
\caption{Displayed above are all available X-ray luminosity observations 
(1995-2008; gray symbols), converted into the common [0.2 - 2.0 keV] energy 
passband ({\it see} \citealt{A09}).  Uncertainties 
($\Delta${\it L}$_{\rm \,X}$) are typically of order $\lesssim$4\% for the 
individual measures.  Also shown is the appropriately modulated and phased 
light curve, as determined utilizing an iterative grid search method, 
incorporating the long-term magnetic activity cycle of $\sim$8.84 years as 
ascertained from all X-ray measures along with the {\it IUE} \ion{C}{4} 
emission and \ion{Mg}{2} Index data.  The vertical range bar near the 
{\it ROSAT} observations at $\sim$JD2450315 demonstrates the extent of the 
effects of rotational modulation of X-ray luminosity due to the presence of 
magnetically active regions, or plages.  Clearly, overall coronal X-ray 
luminosity is more affected by long-term magnetic activity changes than 
by the rotational modulation effects of discretely distributed active regions. 
\label{fig:LongXMM}}
\end{figure}

\begin{figure}
\epsscale{0.7}
\plotone{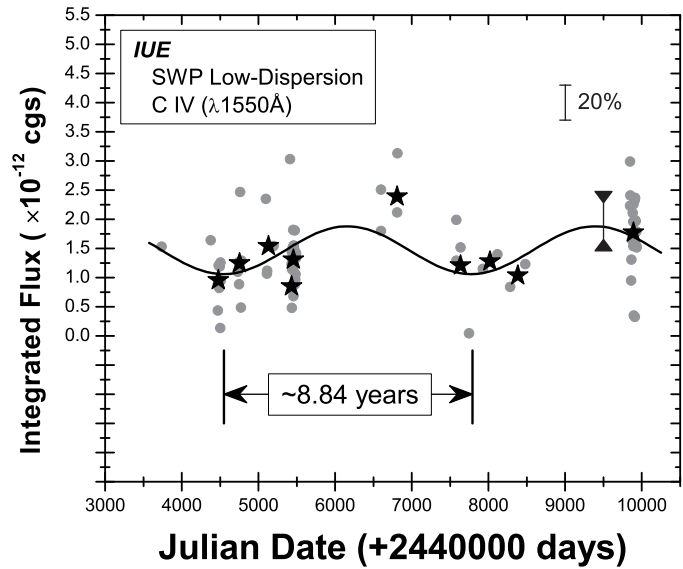}
\caption{Data presented above include all available {\it IUE} \ion{C}{4} 
emission values (1978-1995; gray dots).  The star symbols represent the 
$\sim$yearly averages.  The long-term magnetic activity cycle (properly 
modulated and phased) of $\sim$8.84 years is evident and is determined with 
better than 15\% FDR.  Note that this cycle meshes precisely with the more 
contemporary X-ray observations (see Figure \ref{fig:LongXMM}).  A 20\% 
uncertainty bar, our estimate of the internal consistency of the SWP 
\ion{C}{4} flux levels, is shown in the upper right corner.  The vertical 
range bar near the observations at $\sim$JD2449900 demonstrates the extent of 
rotational modulation on the TR produced \ion{C}{4} emission levels. 
\label{fig:LongCIV}}
\end{figure}

An iterative grid search routine was employed to determine the appropriate 
light modulation amplitude and cycle phase for each of these datasets (X-ray, 
\ion{C}{4}, \ion{Mg}{2}).  Figure \ref{fig:LongXMM} shows the complete X-ray 
luminosity dataset with the best-fitting, 8.84 year period, light curve.  In 
Figure \ref{fig:LongCIV}, the \ion{C}{4} data set is shown with the 
appropriately modulated and phased light curve overplotted.  With regard to 
cycle phasing, no special consideration was necessary to mesh the X-ray and 
{\it IUE} \ion{C}{4} data correctly.  As seen, the earlier \ion{C}{4} 
activity cycling phases directly into the more contemporary X-ray observations.

It must also be noted that contained within these long-term light curves are 
the rotational modulation effects of stellar active regions.  As determined 
below (Section 5.2), the ratio of the long-term magnetic activity cycle 
amplitude to the rotational modulation amplitude ({\it A}$_{\rm \;magnetic}$ / 
{\it A}$_{\rm \;rotation}$) is about 3.9, 0.80, and $\sim$0.5 for the X-ray, 
\ion{C}{4}, and \ion{Mg}{2} observations, respectively.  This implies that 
the X-ray data will be much more sensitive to the long-term magnetic activity 
changes than the rotational effects of active regions.  Conversely, the 
\ion{Mg}{2} Index appears to be the superior choice for more precisely 
determining $\alpha$ Cen B's rotation period, yielding the highest determinacy.

\begin{figure}
\epsscale{0.7}
\plotone{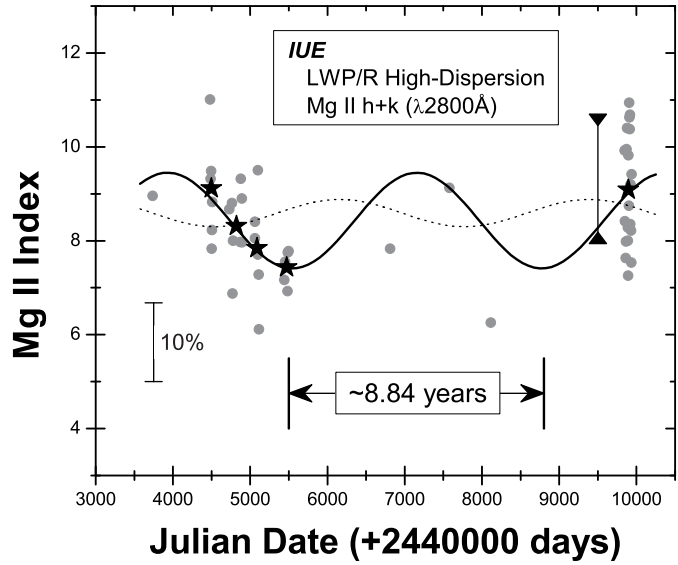}
\caption{Data presented above include all available {\it IUE} \ion{Mg}{2} 
{\it h+k} emission Index values (1978-1995; gray dots).  The star symbols 
represent the $\sim$yearly averages.  Though there is a lack of data in the 
important middle region of the above light curve, a long-term magnetic 
activity cycle of $\sim$8.84 years would seem possible (solid line).  A 10\% 
uncertainty bar, our estimate of the internal consistency of the LWP/R 
\ion{Mg}{2} emission Index values, is shown in the lower left corner.  The 
vertical range bar near the observations at $\sim$JD2449900 demonstrates the 
extent of rotational modulation on the chromospherically-produced \ion{Mg}{2} 
{\it h+k} emission levels.  Unlike the X-ray and \ion{C}{4} observations 
presented above (Figures \ref{fig:LongXMM} and \ref{fig:LongCIV}), the effects 
of rotational modulation clearly dominate the overall emission level.  
Unfortunately, cycle phasing is significantly different to that of the X-ray 
and \ion{C}{4} observations (dashed line).  It is unclear whether this 
measured activity cycle length is physical, but is unexpectedly not 
synchronized with activity levels in the hotter TR and corona, or if the 
randomly sampled rotational effects led to a serendipitous, albeit 
potentially spurious, determination. \label{fig:LongMgII}}
\end{figure}

The long-term chromospheric \ion{Mg}{2} Index dataset is shown in Figure 
\ref{fig:LongMgII}.  Though there is a paucity of data in the important middle 
regions of the timeline, the periodic cycling of $\sim$8.84 years (via 
Lomb-Scargle analysis; FDR $<$ 20\%) would seem possible.  Peculiarly, the 
computed best-fitting cycle phasing is significantly different to that of 
coronal X-ray and TR \ion{C}{4} observations ($\Delta\phi$/2$\pi$ $\approx$ 
0.31, corresponding to a cycle phase lag of $\sim$1000 days!).  An attempt was 
made to ``split the difference'' between the cycle phasing of the \ion{Mg}{2} 
and that of both the \ion{C}{4} and X-ray data sets using the method of 
minimum phase dispersion, but this only resulted in a compromise that visually 
appeared inadequate for all three data sets.

It is possible that since the measured amplitude attributable to rotationally 
modulated \ion{Mg}{2} {\it h+k} emission is significantly larger than the 
variability due to overall long-term chromospheric activity changes, that the 
effects of rotation modulation, combined with the insufficiently sampled 
data is overwhelming the effects attributable to long-term activity, resulting 
in a serendipitously measured $\sim$8.8 year period.  Alternatively, it is 
possible that this measured activity cycle length is indeed real and physical, 
but is simply not synchronized with activity levels in the hotter TR 
(\ion{C}{4}) and corona (X-ray).  Considering the Sun's behavior in which 
the magnetic activity indicators are well-correlated (see, for example, 
\citealt{L97}; \citealt{JSA03}), this latter scenario for $\alpha$ Cen B must 
be considered unlikely. \\

\subsection{Rotation Period, Revisited}

The last {\it IUE} campaign in 1995 provides us with appropriate data sets to 
search for $\alpha$ Cen B's rotation period, with sufficient observations of 
both \ion{Mg}{2} {\it h+k} (20) and \ion{C}{4} (21) over the proper 
timeframe ($\sim$90 days).  These observations were carried out by two of us 
(E.F.G. and L.E.D.) as part of an {\it IUE} GO program.  Additionally, we 
explored potentially viable sections in 1981 (\ion{Mg}{2}) and 1983 
(\ion{C}{4}).

\begin{figure}
\epsscale{0.7}
\plotone{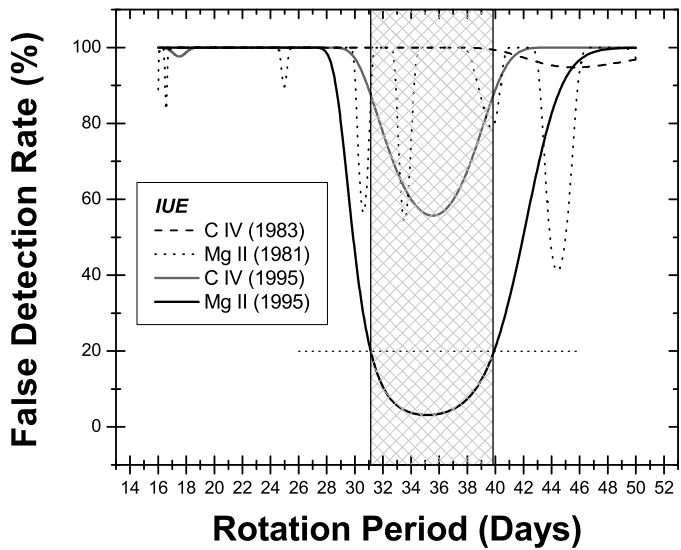}
\caption{Plotted above is the false discovery rate (FDR) as a function of 
short-term light modulation period due to the presence of magnetically active 
regions.  All potentially viable sections of {\it IUE} data are displayed.  
Only the \ion{Mg}{2} Index data obtained in 1995 result in a rotation period 
of high confidence ({\it P}$_{\rm rot}$ $\approx$ 35.1 days; FDR $\approx$ 
5\%), but as seen above, this solution is strongly corroborated by the 1995 
\ion{C}{4} FDR result.  Unfortunately, the other two epochs (1983 \ion{C}{4}; 
1981 \ion{Mg}{2}) were insufficiently sampled in phase space to obtain any 
additional useful rotation modulation information. \label{fig:ShortFDR}}
\end{figure}

\begin{figure}
\epsscale{0.8}
\plotone{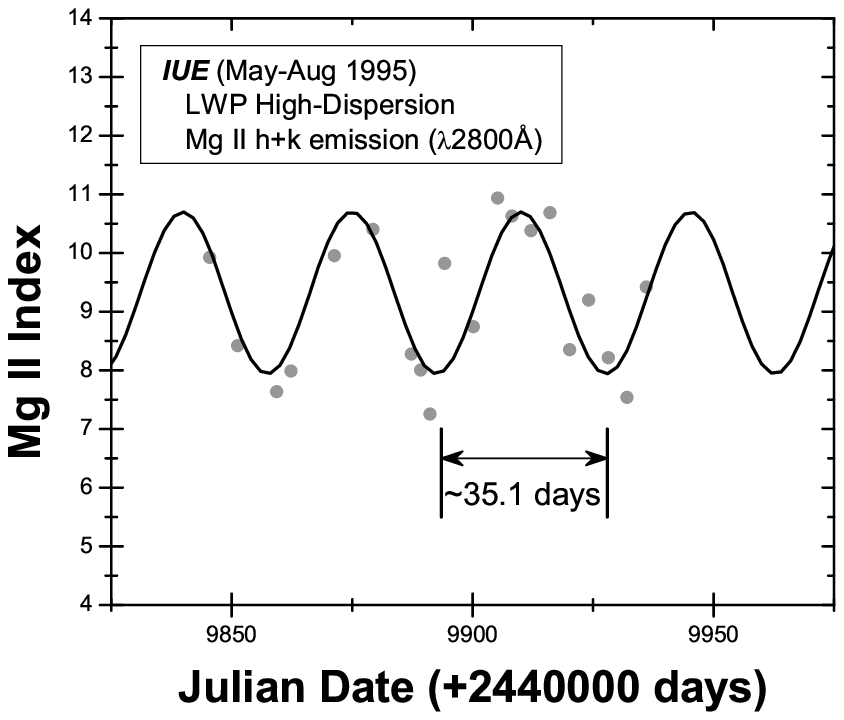}
\caption{1995 {\it IUE} \ion{Mg}{2} Index values (gray dots) shown above 
are overplotted with the appropriately modulated and phased light curve, as 
determined utilizing an iterative grid search method and incorporating the 
$\sim$35.1 day rotation period ascertained from the above (see Figure 
\ref{fig:ShortFDR}) Lomb-Scargle analysis. \label{fig:ShortMgII}}
\end{figure}

\begin{figure}
\epsscale{0.8}
\plotone{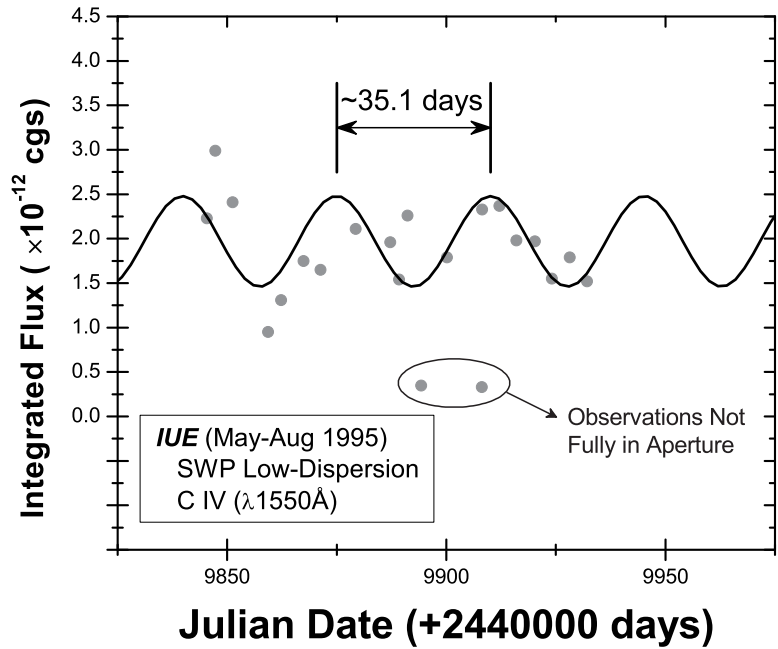}
\caption{As seen above, the determined $\sim$35.1 day rotation rate (properly 
modulated and phased) appears appropriate for the 1995 {\it IUE} \ion{C}{4} 
emission data (gray dots).  Unfortunately, due to the clearly higher degree of 
noise within these observations, high confidence is not, per se, possible. 
\label{fig:ShortCIV}}
\end{figure}

These {\it IUE} datasets were analyzed using the same Lomb-Scargle method 
described above.  In Figure \ref{fig:ShortFDR}, the FDR functions for both 
spectral features observed during the 1995 campaign, along with the two other 
epochs that might have contained, a priori, marginally sufficient data, 
are displayed.  As is apparent, the 1995 \ion{Mg}{2} {\it h+k} Index data 
indicate a {\it P}$_{\rm rotation}$ $\doteq$ 35.1 day rotation period for 
$\alpha$ Cen B with a high level of confidence ($\approx$3\% FDR).  Due to the 
relatively broad nature of the FDR function, periods between about 31.1 and 
39.9 days are plausible at the better than 20\% FDR level.  Though there is a 
higher degree of scatter (noise) within the data, the results of the analysis 
of the 1995 \ion{C}{4} flux variations substantively corroborate this 35.1 
day rotation period, but with an expectedly higher predicted FDR.  Figures 
\ref{fig:ShortMgII} and \ref{fig:ShortCIV} display the 1995 \ion{Mg}{2} 
{\it h+k} Index and \ion{C}{4} emission flux data, respectively, overplotted 
with the appropriately modulated and phased light curve.  Unfortunately, the 
1981 and 1983 data sets were simply not adequately sampled, temporally, to 
derive any exacting rotation modulation information.

Previous spectroscopic determinations of the rotation rate of $\alpha$ Cen B 
have been carried out by Jay et al. (1997; {\it P}$_{\rm rotation}$ = 
36.9 days), and Buccino \& Mauas (2008; 35.1 days).  These two previous 
studies wholly utilize the 1995 {\it IUE} data set(s), but employ different 
data reduction techniques than this study.  Both (thankfully!) arrive at 
essentially the same result as presented here.  With regard to the possibility 
of spot (active region) creation/migration/destruction that may have occurred 
simultaneous to the observations, and/or the effects of differential rotation, 
the small differences between these rotation period values and that of the 
X-ray result above (Section 3.2; 37.8 days) must be considered negligible.  
Without prejudice to author or technique, the average rotation period of 
$\alpha$ Cen B would therefore be {\it P}$_{\rm rotation}$ = 36.23 $\pm$ 1.35 
days.

As insightfully pointed out by the anonymous referee:  \citet{SO97} infer a 
rotation period of 42 days for $\alpha$ Cen B using an empirical relationship 
between the ratio of the chromospheric emission of the Mt. Wilson \ion{Ca}{2} 
HK line cores to total bolometric emission, {\it R}$^{\,\prime}_{\rm HK}$, and 
the Rossby number, {\it R}$_0$, which is itself related to the rotation period 
of the star (see \citealt{NHBDV84}).  The difference ($\sim$14\%) in the 
rotation period determined in this study and the estimate of \citet{SO97} is 
likely in part due to intrinsic scatter in {\it R}$^{\,\prime}_{\rm HK}$ at a 
given {\it R}$_0$ (due to rotational modulation, activity cycles), plus 
problems in calibration of the {\it R}$^{\,\prime}_{\rm HK}$ - {\it R}$_0$ 
relationship, especially at low activity levels \citep{W04,PP04,SGC05,S06}.  
In particular, clear trends with metallicity are not included in the standard 
\citep{NHBDV84} calibration \citep{W04,S06}. $\alpha$ Cen B is rather metal 
rich, which pushes the inferred minimum {\it R}$^{\,\prime}_{\rm HK}$ (and 
likely the whole calibration) to overly low values \citep{S06}. With 
{\it R}$^{\,\prime}_{\rm HK}$ too low (due to uncalibrated high metals), the 
{\it P}$_{\rm rotation}$ derived from {\it R}$^{\,\prime}_{\rm HK}$ would be 
too large, as seen.

\section{STATISTICAL TREATMENT OF PERIOD FDRs}

To further address the robustness of the analytically derived Lomb-Scargle 
FDRs presented throughout this paper, Monte Carlo random permutation 
simulations (see, \citealt{D57}) were carried out.  Random permutation 
tests are restricted to cases where changing the temporal order of the data 
destroys the measured effect.  In our case, power is the test statistic, since 
it is power that leads directly to the FDR value.

The technique involves randomly rearranging the data points, leaving the 
time spacing the same, rerunning the periodogram, and keeping track of the 
number of instances that the resultant power is greater than the original 
value.  There are too many possible orderings to allow complete enumeration 
with the number of data points used in our respective period determinations, 
but fortunately, as pointed out by \citet{D57}, the Monte Carlo simulations 
are asymptotically equivalent to the full permutation test, provided the 
number of iterations is sufficiently large.

\begin{figure}
\epsscale{0.7}
\plotone{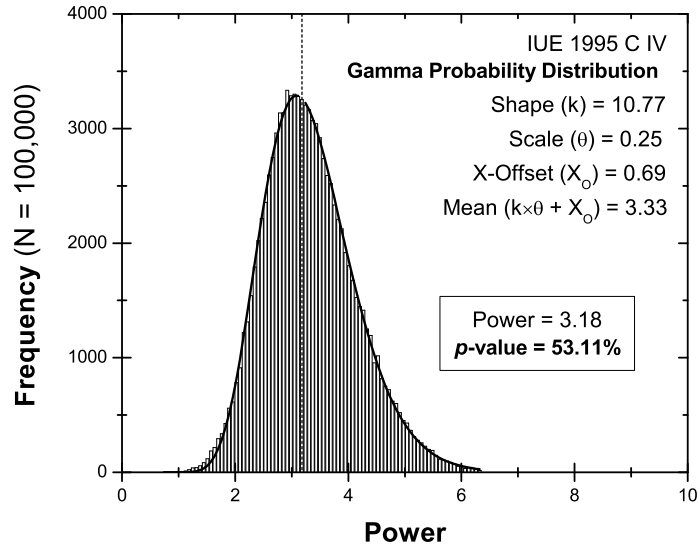}
\caption{Results of the Monte Carlo permutation test for the 1995 {\it IUE} 
\ion{C}{4} data set.  The number of iterations was 100,000, and the resultant 
power frequency is well represented with a Gamma probability distribution 
characterized by the standard shape ({\it k}) and scale ($\theta$) parameters, 
along with a small offset (shift) along the power axis ({\it X}$_0$).  For 
this type of distribution, the mean power is then given by 
{\it k}$\times$$\theta$+{\it X}$_0$.  The statistical {\it p} -value 
represents the fractional number of permutations that result in a greater 
power, and in this case, {\it p} -value = 1 - cdf, where cdf is the 
cumulative distribution function of the resultant Gamma probability 
distribution, evaluated at the desired power (minus {\it X}$_0$).  For this 
dataset, the Lomb-Scargle period analysis results in a rotation rate of 35.1 
days with a power of 3.18 and an analytically derived FDR of 60.5\%.  The 
Monte Carlo permutation analysis shown above implies that the Lomb-Scargle FDR 
is comparable, but overestimates the statistically derived false detection 
rate ({\it p} -value $\approx$ 53.11\%). \label{fig:MCPT}}
\end{figure}

For our tests, the number of iterations was always {\it N} = 100,000.  In all 
cases the resultant power frequency distributions were well represented with 
Gamma probability distributions that could be characterized by the standard 
shape ({\it k}) and scale ($\theta$) parameters, with a small offset (shift) 
along the power axis ({\it X}$_0$).  For these Gamma probability distribution, 
the mean power is then given by {\it k}$\times$$\theta$+{\it X}$_0$.  The 
statistical {\it p}-value represents the fractional number of permutations 
that resulted in a greater power, and correspondingly better FDR.  Therefore,
\begin{eqnarray}
{\it p}{\rm -value} = 1 - {\rm cdf}\,, \nonumber
\end{eqnarray}
\noindent where cdf is the cumulative distribution function appropriate 
to the resultant Gamma probability distribution,  evaluated at the desired 
power (minus {\it X}$_0$).  Shown in Figure \ref{fig:MCPT} is the power 
frequency distribution for the 1995 {\it IUE} \ion{C}{4} data set (used for 
the determination of $\alpha$ Cen B's rotation rate), which incidentally has 
the highest (aka poorest) FDR reported in this study.  As seen, the 
Monte Carlo random permutation test results in a ``true'' statistical FDR of 
$\sim$53\%, compared to the $\sim$61\% reported by Lomb-Scargle.

The results of all of the Monte Carlo simulations are shown in Table 
\ref{tab:STATS} along with the respective Lomb-Scargle FDR values.  Note that 
the analytically determined values given by the Lomb-Scargle method appear to 
be comparable to the values given by the Monte Carlo simulations, but 
consistently overestimate the statistical FDRs.

\section{CONCLUSIONS}

\begin{figure}
\epsscale{0.7}
\plotone{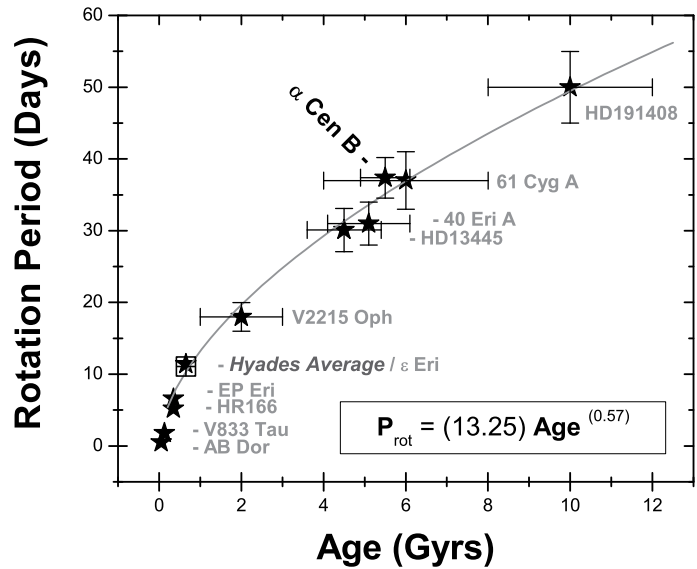}
\caption{Shown above is a plot of rotation period vs. age for a sample of our 
program dK stars.  The box symbol represents the weighted average rotation 
rate value for nine Hyades K0-K5 V stars \citep{RTLDB87}.  Most of all stars 
displayed above have well determined ages and independently measured 
rotation periods (see Table \ref{tab:dKStars}).  The line represents the 
best-fitting power law relation to these data and its functional dependence is 
given in the figure.  Note that the rotation period result 
({\it P}$_{\rm rot}$ = 36.23$\pm$1.35 days) presented in this study for 
$\alpha$ Cen B agrees well with expectations for a typical 5.6 Gyr K1 V star. 
\label{fig:dKRotAge}}
\end{figure}

Independent analyses on the essentially complete sets of X-ray and FUV 
observations, along with the extensive {\it IUE} UV/NUV data sets, have 
yielded determinations of the long-term magnetic activity cycle length 
({\it P}$_{\rm cycle}$ = 8.84$\pm$0.4 yr) and the rotation period 
({\it P}$_{\rm rotation}$ = 36.2$\pm$1.4 days) of the nearby K dwarf, $\alpha$ 
Cen B.  Figure \ref{fig:dKRotAge} places $\alpha$ Cen B in context with the 
other stars in our dK star program (see also Table \ref{tab:dKStars}).  
We note that the rotation period result determined within this study agrees 
well with expectations for a typical 5.6 Gyr K1 V star.

\acknowledgments

Since our {\it FUSE} Cycles 7 and 8 observations occurred after the failure of 
two reaction wheels in 2001 December, there were some additional 
``difficulties'' for {\it FUSE} pointing and acquisition.  Fortunately (all 
credit is due to the heroic efforts of the {\it FUSE} team -- battling roll 
angles, drift, and jitter), the individual spectra of $\alpha$ Cen A and B 
were ultimately secured.  B-G Andersson was also of great individual help in 
the preparation, acquisition, and analysis stages.  We also thank Tom Woods 
({\it LASP}, UC Boulder) for his assistance and expertise with the extraction 
and interpretation of the {\it Solar EUV Experiment} data used herein.  In 
addition, we gratefully acknowledge comments and observations made by the 
anonymous referee and support by NASA Grants NNX06AC45G, NNX08AG95G, and the 
Villanova University Research for Undergraduates Award Program.



\clearpage

\clearpage

\clearpage

\clearpage

\clearpage

\clearpage

\clearpage

\clearpage

\clearpage

\clearpage

\clearpage

\begin{turnpage}

\begin{deluxetable}{lrccccccl}
\tablewidth{0pc}
\tablecaption{Properties of Representative K-type Program Stars \label{tab:dKStars}}
\startdata
\tableline
\tableline
Object                      & HD           & Sp.        & $V$        & $T_{\rm eff}$ & Dist.        & $P_{\rm rot}$         & Age               & Age                                                 \\
Name                        & Number       & Type       & (mag)      & (K)           & (pc)         & (days)                &(Gyr)              & Indicator                                           \\
\tableline
Sun~$^{1}$                  & \nodata      & G2~V       & -26.74     & 5777          & 1~{\rm AU}   & 25.38                 & ~4.6~             & Isotopic Dating                                     \\
\tableline
AB Dor~$^{1}$               & 36705        & K1~V       & 6.93       & 5260          & 14.9~~       & ~~0.514               & ~0.05             & AB Dor Moving Group                                 \\
V833 Tau                    & 283750       & K2~V       & 8.42       & 4600          & 17.9~~       & ~~1.81~               & ~0.13             & Pleiades Str                                        \\
HR 08                       & 166          & K0~V       & 6.13       & 5400          & 13.7~~       & ~~5.2~~               & ~0.35             & UMa Str                                             \\
EP Eri                      & 17925        & K1~V       & 6.03       & 5100          & 10.4~~       & ~~6.6~~               & ~0.35             & UMa Str                                             \\
$\epsilon$ Eri~$^{1}$       & 22049        & K2~V       & 3.73       & 5050          & ~3.22~       & ~11.4~~               & ~0.65             & Hyades Str                                          \\
{\it Hyades Average}~$^{2}$ & \nodata      & \nodata    & \nodata    & \nodata       & ~\nodata~~   & $<$11.07$\pm$1.54$>$  & ~0.65             & Hyades Str                                          \\
V2215 Oph                   & 156026       & K5~V       & 6.34       & 4480          & ~5.4~~       & $\sim$18~~~           & $\sim$2.0~        & Activity/Rotation Relation                          \\
40 Eri A                    & 26965        & K1~V       & 4.43       & 5300          & ~5.04~       & ~31$\pm$3             & ~5.1~             & Age of WD Companion                                 \\
HR 637~$^{1}$               & 13445        & K1~V       & 6.12       & 5240          & 10.9~~       & ~30.1~~               & ~4.5~             & Age-Activity Relation                               \\
{\bf $\alpha$ Cen B}        & {\bf 128621} & {\bf K1~V} & {\bf 1.33} & {\bf 5316}    & ~{\bf 1.347} & {\bf 36.23$\pm$1.35}  & {\bf 5.6$\pm$0.6} & {\bf Isochrones of $\alpha$ Cen A (\citealt{FA78})} \\
61 Cyg A                    & 201091       & K5~V       & 5.22       & 4450          & ~3.4~~       & $\sim$37~~~           & $\sim$6.0$\pm$2   & 61 Cyg Moving Group (\citealt{E96})                 \\
HR 7703                     & 191408       & K3~V       & 5.31       & 4893          & ~6.05~       & $\sim$50~~~           & ~10.0$\pm$2       & HVel star (Old Disk)                                \\
\enddata
\tablenotetext{~}{$^{1}$ Star with planet/brown dwarf companion(s):  AB Dor $\approx$ 93 $M_{J}$; $\epsilon$ Eri $\approx$ 0.92 $M_{J}$; HD 13445 $\approx$ 4.2 $M_{J}$.}
\tablenotetext{~}{$^{2}$ Weighted average rotation period value derived from nine K0-K5 V directly measured Hyades members \citep{RTLDB87}.}
\end{deluxetable}

\end{turnpage}

\clearpage

\begin{deluxetable}{lcccc}
\vspace{5mm}
\tablewidth{0pc}
\tablecaption{The $\alpha$ Centauri System \label{tab:aCenProp}}
\startdata
\tableline
\tableline
Property                                     & Sun          & $\alpha$ Cen A & $\alpha$ Cen B & $\alpha$ Cen C \\
                                             & (Sol)        & HD 128620      & HD 128621      & (Proxima Cen)  \\
\tableline
Spectral type                                & G2 V         & G2 V           & K1 V           & dM5e           \\
{\it V}~\tablenotemark{a}                    & -26.74       & -0.01          & 1.33           & 11.05          \\
{\it (B-V)}~\tablenotemark{a}                & 0.648        & 0.71           & 0.88           & 1.97           \\
{\it (b-y)}~\tablenotemark{b}                & 0.403        & 0.414          & 0.524          & \nodata        \\
Temperature (K)~\tablenotemark{c}            & 5779         & 5847           & 5316           & $\sim$3050     \\
$[$Fe/H$]$~\tablenotemark{d}                 & $\equiv$0.0  & 0.25           & 0.25           & (0.25)         \\
Mass ({\it M}$_{\odot}$)~\tablenotemark{e}   & $\equiv$1.00 & 1.09           & 0.90           & 0.123          \\
Radius ({\it R}$_{\odot}$)~\tablenotemark{f} & $\equiv$1.00 & 1.22           & 0.86           & 0.145          \\
Age (Gyr)~\tablenotemark{g}                  & 4.58         & $\sim$5.6      & $\sim$5.6      & $\sim$5.6      \\
{\it P}$_{\rm rot}$ (days)~\tablenotemark{h} & 25.38        & 15 - 20        & 36.23$\pm$1.35 & $\sim$83.1     \\
\enddata
\tablenotetext{a}{~\citealt{HJ82}.}
\tablenotetext{b}{~\citealt{CB70} (A+B), \citealt{E78b} (A,B).}
\tablenotetext{c}{~\citealt{PLK08}, cf \citealt{NM97,MPLTB00} (A, B), and cf \citealt{SKFQ03} (C).}
\tablenotetext{d}{~\citealt{PLK08} (A, B); \citealt{P80} (C).}
\tablenotetext{e}{~See \citealt{E78a,H81,DGA86,AOP94,PNMBTMJPCBPHSRK02,Y08} (A, B), cf \citealt{SKFQ03} (C).}
\tablenotetext{f}{~\citealt{KTSBLMP03} (A, B), \citealt{SKFQ03} (C).}
\tablenotetext{g}{~See \citealt{FA78,DGA86,GD00,TPMBBC02,ECTMMCB04,Y07,Y08}.}
\tablenotetext{h}{~Equatorial (Sun), L.E. DeWarf, (2010, in preparation) (A), this study (B), \citealt{EGM09} (C).}
\end{deluxetable}


\begin{deluxetable}{ccccc}
\vspace{5mm}
\tablewidth{0pc}
\tablecaption{Published Estimates of the Sun's Long-term Cyclic Variations in X-ray Luminosity (see \citealt{JSA03}) \label{tab:XSun}}
\startdata
\tableline
\tableline
Bandpass    & Log {\it L}$_{\rm X,min}$ & Log {\it L}$_{\rm X,max}$ & {\it L}$_{\rm max}$/{\it L}$_{\rm min}$ & Ref \\
(keV)       & (erg s$^{-1}$           & (erg s$^{-1}$)              & Ratio                                   &     \\
\tableline
0.15 - ~4.0 & 26.0                    & 27.4                        & 25.1                                    & 1~~ \\
0.15 - ~4.0 & 27.0                    & 28.0                        & 10.0                                    & 2,3 \\
0.15 - ~4.0 & 27.2                    & 27.8                        & ~4.0                                    & 4,5 \\
0.04 - 12.4 & 25.5                    & 27.0                        & 31.6                                    & 6~~ \\
0.1~ - ~4.0 & 25.6                    & 27.1                        & 32.6                                    & 7~~ \\
0.1~ - ~2.4 & 26.7                    & 27.3                        & ~4.0                                    & 8~~ \\
0.2~ - ~2.0 & 26.8                    & 27.4                        & ~4.0                                    & 9~~ \\
0.1~ - ~2.4 & 26.0                    & 27.7                        & 50.0                                    & 10~ \\
0.1~ - ~2.4 & 26.8                    & 27.9                        & 12.6                                    & 11~ \\
\enddata
\tablenotetext{~}{~References:  (1) \citealt{PGRV81}; (2) \citealt{VR78}; (3) \citealt{GHPRV82}; 
(4) \citealt{S83}; (5) \citealt{RS87}; (6) cf \citealt{HS96}; (7) \citealt{A96}; 
(8) \citealt{S97}; (9) \citealt{A97}; (10) \citealt{OPR01}; (11) \citealt{JSA03}.}
\end{deluxetable}


\begin{deluxetable}{lcccc}
\vspace{5mm}
\tablewidth{0pc}
\tablecaption{{\it FUSE} Observing Log \label{tab:FUSELog}}
\startdata
\tableline
\tableline
Object           & HD        & Observation  & Aperture & Exposure \\
Name             & Number    & Date         &          & (s)    \\
\tableline
$\alpha$ Cen A   & HD 128620 & 2001 Jun 25  &   MDRS   &   15,332 \\
$\alpha$ Cen B   & HD 128621 & 2001 Jun 24  &   MDRS   &   22,742 \\
\hline
$\alpha$ Cen A   & HD 128620 & 2006 May 5   &   MDRS   &   12,606 \\
$\alpha$ Cen B   & HD 128621 & 2006 May 6   &   MDRS   &   ~~9245 \\
$\alpha$ Cen A+B & \nodata   & 2006 May 4   &   LWRS   &   ~~5316 \\
\hline
$\alpha$ Cen A   & HD 128620 & 2007 Jun 18  &   MDRS   &   13,289 \\
$\alpha$ Cen B   & HD 128621 & 2007 Jun 19  &   MDRS   &   ~~3565 \\
\enddata
\end{deluxetable}

\clearpage

\begin{deluxetable}{lccccc}
\vspace{3cm}
\tablewidth{0pc}
\tablecaption{{\it FUSE} FUV Integrated Emission Fluxes \label{tab:FUVFlux}}
\startdata
\tableline
\tableline
Atomic                         & Wavelength     & Integration Range & \multicolumn{3}{c}{Integrated Emission Flux~\tablenotemark{c}}            \\
Species                        & (\AA)          & (\AA)             & 2001.45                     & 2006.34                     & 2007.47       \\
\tableline
C {\smc ii}~\tablenotemark{a}  &       1036+7   & 1036.00 - 1037.25 & 1.17$\pm$0.68             & 0.49$\pm$\hspace{-1mm}\nodata & 0.33$\pm$0.02 \\
C {\smc iii}                   &        977.020 & ~976.50 - ~977.50 & 4.12$\pm$0.34             & 1.70$\pm$0.37                 & 2.09$\pm$0.68 \\
C {\smc iii}~\tablenotemark{b} & $\sim$1176     & 1174.50 - 1176.75 & 2.98$\pm$0.60             & 1.25$\pm$0.53                 & 1.11$\pm$0.03 \\
O {\smc vi}                    &       1031.925 & 1031.50 - 1032.50 & 4.15$\pm$0.76             & 1.47$\pm$0.49                 & 1.41$\pm$0.29 \\
O {\smc vi}                    &       1037.614 & 1037.25 - 1038.00 & 2.13$\pm$0.33             & 0.63$\pm$0.34                 & 0.68$\pm$0.22 \\
\enddata
\tablenotetext{a}{~Individually combined flux values for the C~{\sc ii} doublet at 1036.337 and 1037.018\AA.}
\tablenotetext{b}{~Blended fluxes from C~{\sc iii} multiplet (six transitions) around 1176\AA.}
\tablenotetext{c}{~Integrated emission flux units:  ($\times$ 10$^{-11}$ erg cm$^{-2}$ s$^{-1}$).}
\end{deluxetable}


\begin{deluxetable}{lccccc}
\vspace{3cm}
\tablewidth{0pc}
\tablecaption{{\it FUSE} FUV Integrated Emission Flux Ratios \label{tab:FluxRatio}}
\startdata
\tableline
\tableline
Atomic                         & Wavelength     & Temperature  & Stellar         & \multicolumn{2}{c}{Integrated Emission Flux Ratios} \\
Species                        & (\AA)          & (K)          & Region          & (2001.45)/(2006.34)   & (2001.45)/(2007.47)         \\
\tableline
C~{\smc ii}~\tablenotemark{a}  &       1036+7   &  $\sim$20000 & Chromosphere/TR & 2.36$\pm$\hspace{-1mm}\nodata  & 3.57$\pm$2.07      \\
C~{\smc iii}                   &        977.020 &  $\sim$50000 & TR              & 2.43$\pm$0.56                  & 1.98$\pm$0.66      \\
C~{\smc iii}~\tablenotemark{b} & $\sim$1176     &  $\sim$50000 & TR              & 2.38$\pm$1.12                  & 2.68$\pm$0.55      \\
O~{\smc vi}                    &       1031.925 & $\sim$300000 & TR              & 2.83$\pm$1.07                  & 2.92$\pm$0.81      \\
O~{\smc vi}                    &       1037.614 & $\sim$300000 & TR              & 3.37$\pm$1.90                  & 3.12$\pm$1.12      \\
\enddata
\tablenotetext{a}{~Individually combined fluxes from C~{\sc ii} doublet at 1036.337 and 1037.018\AA.}
\tablenotetext{b}{~Blended fluxes from C~{\sc iii} multiplet (six transitions) around 1176\AA.}
\end{deluxetable}

\clearpage

\begin{turnpage}

\begin{deluxetable}{ccccccccc}
\tablewidth{0pc}
\tablecaption{{\it IUE} {\smc C iv} Integrated Emission Line Flux Values \label{tab:IUESWP}}
\startdata
\tableline
\tableline
SWP LL $^{a}$ & Julian Date     & Integrated CIV Flux      & SWP LL $^{a}$ & Julian Date     & Integrated CIV Flux      & SWP LL $^{a}$ & Julian Date     & Integrated CIV Flux      \\
Seq. No.      & (+2440000 days) & ($\times$10$^{-12}$ cgs) & Seq. No.      & (+2440000 days) & ($\times$10$^{-12}$ cgs) & Seq. No.      & (+2440000 days) & ($\times$10$^{-12}$ cgs) \\
\tableline
SWP02320      & 3737.959        & 1.53                     & SWP19772      & 5445.297        & 1.51                     & SWP38115      & 7925.409        & 1.15                     \\
SWP09035      & 4377.873        & 1.64                     & SWP19837      & 5452.462        & 1.82                     & SWP39442      & 8115.062        & 1.40                     \\
SWP09818      & 4468.679        & 0.44                     & SWP19839      & 5452.522        & 1.13                     & SWP40690      & 8283.598        & 0.84                     \\
SWP09928      & 4479.767        & 1.21                     & SWP19882      & 5457.317        & 0.68                     & SWP42233      & 8481.140        & 1.23                     \\
SWP10016      & 4488.001        & 0.83                     & SWP19885      & 5457.411        & 1.54                     & SWP54631      & 9845.357        & 2.23                     \\
SWP10093      & 4495.008        & 1.19                     & SWP19886      & 5457.437        & 1.32                     & SWP54647      & 9847.250        & 2.99                     \\
SWP10164      & 4500.007        & 0.14                     & SWP19912      & 5461.205        & 1.55                     & SWP54673      & 9851.222        & 2.41                     \\
SWP10193      & 4503.913        & 1.26                     & SWP19915      & 5461.315        & 1.53                     & SWP54724      & 9859.348        & 0.95                     \\
SWP10214      & 4507.016        & 0.95                     & SWP19950      & 5465.213        & 1.49                     & SWP54744      & 9862.267        & 1.31                     \\
SWP13887      & 4729.243        & 1.10                     & SWP19953      & 5465.326        & 1.34                     & SWP54793      & 9867.384        & 1.75                     \\
SWP13988      & 4742.330        & 1.27                     & SWP20015      & 5472.325        & 0.88                     & SWP54844      & 9871.317        & 1.65                     \\
SWP14038      & 4748.399        & 0.89                     & SWP20018      & 5472.429        & 1.81                     & SWP54975      & 9879.303        & 2.11                     \\
SWP14202      & 4762.165        & 2.47                     & SWP20019      & 5472.456        & 1.37                     & SWP55042      & 9887.234        & 1.96                     \\
SWP14256      & 4770.169        & 0.49                     & SWP20064      & 5479.397        & 1.34                     & SWP55051      & 9889.205        & 1.54                     \\
SWP14302      & 4778.165        & 1.29                     & SWP20066      & 5479.468        & 0.97                     & SWP55060      & 9891.155        & 2.26                     \\
SWP16922      & 5097.375        & 2.35                     & SWP20088      & 5483.311        & 1.06                     & SWP55114      & 9894.225        & 0.35                     \\
SWP16997      & 5109.428        & 1.07                     & SWP20140      & 5489.344        & 1.42                     & SWP55178      & 9900.169        & 1.79                     \\
SWP17027      & 5113.416        & 1.12                     & SWP20141      & 5489.371        & 1.34                     & SWP55233      & 9908.106        & 0.33                     \\
SWP17697      & 5197.129        & 1.54                     & SWP28488      & 6596.263        & 2.51                     & SWP55234      & 9908.166        & 2.33                     \\
SWP19478      & 5411.515        & 3.03                     & SWP28489      & 6596.291        & 1.80                     & SWP55269      & 9912.140        & 2.37                     \\
SWP19663      & 5431.350        & 0.86                     & SWP30092      & 6809.515        & 2.12                     & SWP55307      & 9915.995        & 1.98                     \\
SWP19664      & 5431.378        & 1.23                     & SWP30100      & 6810.569        & 3.13                     & SWP55338      & 9920.172        & 1.97                     \\
SWP19702      & 5435.459        & 0.48                     & SWP35607      & 7579.908        & 1.99                     & SWP55357      & 9924.103        & 1.55                     \\
SWP19705      & 5435.543        & 1.44                     & SWP35608      & 7579.934        & 1.29                     & SWP55377      & 9928.172        & 1.79                     \\
SWP19727      & 5439.473        & 1.26                     & SWP36037      & 7636.431        & 1.52                     & SWP55405      & 9932.069        & 1.52                     \\
SWP19729      & 5439.536        & 1.14                     & SWP36808      & 7746.006        & 0.05                     &               &                 &                          \\
\enddata
\tablenotetext{a}{~{\it IUE} short wavelength primary camera, low dispersion, large aperture.}
\end{deluxetable}

\end{turnpage}

\clearpage

\begin{turnpage}

\begin{deluxetable}{ccccccccc}
\tablewidth{0pc}
\tablecaption{{\it IUE} {\smc Mg ii} h+k Emission Indexes \label{tab:IUELWP}}
\startdata
\tableline
\tableline
LWP/R HL $^{a}$ & Julian Date     & Mg II {\it h+k} & LWP/R HL $^{a}$ & Julian Date     & Mg II {\it h+k} & LWP/R HL $^{a}$ & Julian Date     & Mg II {\it h+k} \\
Seq. No.        & (+2440000 days) & Index           & Seq. No.        & (+2440000 days) & Index           & Seq. No.        & (+2440000 days) & Index           \\
\tableline
LWR02095        & 3737.873        &                 & LWR12908        & 5059.353        &                 & LWP30740        & 9859.341        & ~7.63           \\
LWR02096        & 3737.898        & ~8.96           & LWR12909        & 5059.384        &                 & LWP30763        & 9862.311        & ~7.99           \\
LWR02097        & 3737.919        &                 & LWR13193        & 5097.379        & 7.71            & LWP30835        & 9871.311        & ~9.96           \\
LWR08526        & 4468.691        &                 & LWR13220        & 5100.398        & 9.50            & LWP30863        & 9879.299        & 10.40           \\
LWR08640        & 4479.760        & 11.01           & LWR13221        & 5100.428        &                 & LWP30916        & 9887.261        & ~8.28           \\
LWR08722        & 4487.996        & ~9.32           & LWR13222        & 5100.463        &                 & LWP30930        & 9889.233        & ~8.00           \\
LWR08778        & 4495.006        & ~9.49           & LWR13277        & 5109.425        & 7.28            & LWP30940        & 9891.134        & ~7.26           \\
LWR08830        & 4500.004        & ~7.83           & LWR13303        & 5113.413        & 6.12            & LWP30946        & 9894.202        & ~9.82           \\
LWR08858        & 4503.910        & ~8.23           & LWR13958        & 5197.127        &                 & LWP30962        & 9900.163        & ~8.75           \\
LWR08884        & 4507.013        & ~8.83           & LWR15736        & 5439.566        & 7.17            & LWP31015        & 9905.220        & 10.94           \\
LWR10517        & 4729.239        & ~8.67           & LWR15826        & 5452.553        & 7.55            & LWP31038        & 9908.159        & 10.63           \\
LWR10795        & 4762.162        & ~8.81           & LWR16023        & 5483.308        & 6.93            & LWP31078        & 9912.134        & 10.38           \\
LWR10854        & 4770.166        & ~6.88           & LWR16049        & 5486.227        & 7.75            & LWP31093        & 9916.022        & 10.69           \\
LWR10928        & 4778.162        & ~8.00           & LWR16120        & 5496.172        & 7.78            & LWP31139        & 9920.166        & ~8.35           \\
LWR11603        & 4870.114        & ~7.98           & LWP09929        & 6809.583        & 7.83            & LWP31161        & 9924.096        & ~9.20           \\
LWR11643        & 4878.035        & ~9.32           & LWP15073        & 7579.762        & 9.13            & LWP31179        & 9928.164        & ~8.22           \\
LWR11762        & 4890.068        & ~8.90           & LWP15074        & 7579.795        &                 & LWP31206        & 9932.061        & ~7.54           \\
LWR11803        & 4896.049        & ~7.97           & LWP18561        & 8114.983        & 6.26            & LWP31240        & 9936.058        & ~9.42           \\
LWR12906        & 5059.299        & ~8.41           & LWP30640        & 9845.381        & 9.93            &                 &                 &                 \\
LWR12907        & 5059.328        & ~8.06           & LWP30691        & 9851.198        & 8.42            &                 &                 &                 \\
\enddata
\tablenotetext{a}{~{\it IUE} long wavelength primary/redundant camera, high dispersion, large aperture.}
\end{deluxetable}

\end{turnpage}

\clearpage

\begin{deluxetable}{lcccccccc}
\vspace{3cm}
\tablewidth{0pc}
\tablecaption{Statistical Treatment of False Detection Rates \label{tab:STATS}}
\startdata
\tableline
\tableline
                &               & \multicolumn{2}{c}{Lomb-Scargle} & \multicolumn{5}{c}{Monte Carlo Permutation Test}                                         \\
Data Set        & Period $^{a}$ & Power $^{b}$     & FDR $^{c}$    & {\it k} (shape) & $\theta$ (scale) & {\it X}$_0$ & Mean $^{d}$ & {\it p}-value $^{e}$    \\
\tableline
\multicolumn{8}{l}{Long-term magnetic activity cycle}                                                                                                         \\
X-ray           & 3230          & 16.90            & $\approx$0    & ~5.96           & 0.53             & 0.28               & 3.44        & $\approx$0       \\
C {\smc iv}     & 3230          & ~7.75            & ~3.8          & ~5.17           & 0.62             & 0.46               & 3.69        & ~1.14            \\
Mg {\smc ii}    & 3230          & ~5.96            & 13.0          & ~4.89           & 0.63             & 0.14               & 3.24        & ~4.44            \\
\multicolumn{8}{l}{Rotation modulation}                                                                                                                       \\
X-ray           & 37.8          & 10.11            & ~0.1          & ~5.88           & 0.40             & 1.21               & 3.59        & $\approx$0       \\
C {\smc iv}     & 35.1          & ~3.18            & 60.5          & 10.77           & 0.25             & 0.69               & 3.33        & 53.11            \\
Mg {\smc ii}    & 35.1          & ~6.28            & ~4.9          & ~6.16           & 0.38             & 1.15               & 3.47        & ~0.85            \\
\enddata
\tablenotetext{a}{~Reported period value in days.}
\tablenotetext{b}{~Measured Lomb-Scargle power of the reported period value.}
\tablenotetext{c}{~Analytically-derived false detection rate determined by the Lomb-Scargle analysis.}
\tablenotetext{d}{~For the resultant Gamma probability distributions, the mean power = {\it k}$\times$$\theta$ + {\it X}$_O$.}
\tablenotetext{e}{~Fractional number of permutations that result in a higher measured power.  In this case, {\it p}-value = 1-cdf, 
were cdf is the cumulative distribution function of the resultant Gamma probability distribution, evaluated at the measured power, offset by {\it -X}$_O$.}
\end{deluxetable}

\end{document}